\documentclass[sigconf]{acmart}

\AtBeginDocument{%
  }

\usepackage{acmart-taps}

\usepackage{algorithmic}
\usepackage{calc}
\usepackage[inline,shortlabels]{enumitem}
\usepackage[utf8x]{inputenc} 
\usepackage{lscape}
\usepackage{listings}
\usepackage{makecell}
\usepackage{multirow}
\usepackage{stfloats} 
\usepackage{textcomp}
\usepackage{xspace}

\newif\ifrevision
\newif\ifroundOne

\revisionfalse
\roundOnefalse

\ifrevision
    \ifroundOne
        \definecolor{rblue}{HTML}{3F91A0}
        \newcommand{\new}{\color{rblue}}   % Start blue text
        \newcommand{\old}{}   % Return to black
        \newcommand{\del}[1]{\color{red}{#1}}
        \newcommand{\newR}[1]{\color{rblue}{}}
        \newcommand{\delR}[1]{\color{red}{}}
    \else
        \newcommand{\new}{}
        \newcommand{\old}{}
        \newcommand{\del}[1]{}    
        \newcommand{\newR}[1]{\color{blue}{#1}}
        \newcommand{\delR}[1]{\color{orange}{#1}}
    \fi
\else
    \newcommand{\new}{}
    \newcommand{\old}{}
    \newcommand{\del}[1]{}            % Deletes the argument entirely
    \newcommand{\newR}{}
    \newcommand{\delR}[1]{}
\fi

\newcommand{\ie}{\emph{i.e.\@}\xspace}
\newcommand{\eg}{\emph{e.g.\@}\xspace}

\newcommand{\mypara}[1]{\smallskip\noindent\textbf{\textit{#1.}}\,\xspace}

\newcommand{\spara}[1]{\smallskip\textit{#1.}\,\xspace}

\newcommand{\scode}[1]{{\texttt{\small \detokenize{#1}}}}
\newcommand{\scodewp}[1]{{\texttt{\small #1}}}

\newcommand{\sysraw}{Tidynote}
\newcommand{\titletext}{\sysraw: Always-Clear \del{Computational Notebooks}\new{Notebook Authoring}\old}

\newcommand{\sys}{\textsc{\sysraw}\xspace}

\newcommand{\spad}{scratchpad\xspace}
\newcommand{\Spad}{Scratchpad\xspace}

\aptLtoXcmd{
\long\def\circledletter#1{\xbox{aptbox}{\XMLaddatt{style}{border: 0px solid black;background-color: \#42a5f5;color:\#FFFFFF;border-radius: 50\%; padding: 0.04em 0.3em;text-align:center;}{#1}}}}{}

\DeclareRobustCommand*{\circledletter}[1]{\raisebox{-2pt}{
\hspace{-.5em}
\includegraphics[width=9.3pt]{images/#1}
\hspace{-.5em}
\Description{A circle with letter #1}
}}

\copyrightyear{2026}
\acmYear{2026}
\setcopyright{cc}
\setcctype{by}
\acmConference[CHI '26]{Proceedings of the 2026 CHI Conference on Human Factors in Computing Systems}{April 13--17, 2026}{Barcelona, Spain}
\acmBooktitle{Proceedings of the 2026 CHI Conference on Human Factors in Computing Systems (CHI '26), April 13--17, 2026, Barcelona, Spain}
\acmPrice{}
\acmDOI{10.1145/3772318.3791919}
\acmISBN{979-8-4007-2278-3/2026/04}

\newcommand{\DGOne}{DG1: Providing \new flexible \old
structures for exploratory workflows\xspace}
\newcommand{\DGTwo}{DG2: Enabling rapid iteration between exploratory and non-exploratory activities\xspace}
\newcommand{\DGThree}{DG3: Promoting clarity in program state\xspace}

\newcommand{\RQ}[1]{\textbf{RQ{#1}}}

\newcommand{\tidinessEvalCaption}[0]{
    \newR{Evaluation of the final notebooks created by participants in the study.
    Each code statement was categorized as: (1) \emph{Relevant} (R) if directly generating results for the analysis tasks; (2) \emph{Necessary} (N) if not relevant but syntactically necessary for the code to run; (3) \emph{Transient} (T) if neither relevant nor necessary, such as one-off exploration.
    }
}

\newcommand{\customnormal}{\fontsize{9pt}{11pt}\selectfont}
\newcommand{\customheader}{\fontsize{8.5pt}{10pt}\selectfont}

\newcommand{\tidinessEvalDescription}[0]{A table comparing code statement classification counts for 13 participants across two Tidynote containers (as column groups): Notebook and Scratchpad. Each column group contains counts for Relevant, Necessary, and Transient code statements. The Notebook column group shows larger counts for Relevant and Necessary code than Transient code across all participants, on average 8.6 Relevant statements, 10.5 Necessary statements, and 0.8 Transient statements. The Scratchpad column group shows zeros for Relevant and Necessary code across all participants. The bottom rows provide Minimum, Mean, and Maximum aggregate statistics for all columns.}

\lstdefinelanguage{TypeScript}{
  morekeywords={
    abstract, any, as, async, await, boolean, break, case, catch, class, const, continue,
    debugger, declare, default, delete, do, else, enum, export, extends, false, finally,
    for, from, function, get, if, implements, import, in, infer, instanceof, interface,
    is, keyof, let, module, namespace, never, new, null, number, object, of, package,
    private, protected, public, readonly, require, return, set, static, string, super,
    switch, symbol, this, throw, true, try, type, typeof, undefined, unknown, var, void,
    while, with, yield
  },
  sensitive=true,
  morecomment=[l]{//},
  morecomment=[s]{/*}{*/},
  morestring=[b]",
  morestring=[b]',
  morestring=[b]`,
  keywordstyle=\color{blue}\bfseries,
  commentstyle=\color{gray}\ttfamily,
  stringstyle=\color{red}\ttfamily,
  basicstyle=\ttfamily\footnotesize,
  breaklines=true,
  showstringspaces=false
}

\begin{document}

\title{\titletext}

\author{Ruanqianqian (Lisa) Huang}
\orcid{0000-0002-4242-419X}
\affiliation{\institution{University of California San Diego} \city{La Jolla} \state{CA} \country{USA}}
\email{r6huang@ucsd.edu}

\author{Brian Hempel}
\orcid{0000-0003-3466-5556}
\affiliation{\institution{University of California San Diego} \city{La Jolla} \state{CA} \country{USA}}
\email{bhempel@ucsd.edu}

\author{Yining Cao}
\orcid{0000-0002-3962-2830}
\affiliation{\institution{University of California San Diego} \city{La Jolla} \state{CA}   \country{USA}}
\email{rimacyn@ucsd.edu}

\author{James Hollan}
\orcid{0000-0001-9568-2646}
\affiliation{\institution{University of California San Diego} \city{La Jolla} \state{CA} \country{USA}}
\email{hollan@ucsd.edu}

\author{Haijun Xia}
\orcid{0000-0002-9425-0881}
\affiliation{\institution{University of California San Diego} \city{La Jolla} \state{CA} \country{USA}}
\email{haijunxia@ucsd.edu}

\author{Sorin Lerner}
\orcid{0000-0003-3957-0628}
\affiliation{\institution{Cornell University} \city{Ithaca} \state{NY}\country{USA}}
\email{sorin.lerner@cornell.edu}

\renewcommand{\shortauthors}{Huang et al.}

\begin{abstract}
Recent work identified \emph{clarity} as one of the top quality attributes that notebook users value, but notebooks lack support for maintaining clarity throughout the exploratory phases of the notebook authoring workflow.
We propose \emph{always-clear\new{ notebook authoring}} \del{computational notebooks}\old that supports both clarity and exploration, and present \del{a Jupyter prototype called \sys that supports always-clear authoring}\new{a Jupyter implementation called \sys}. \old
The key to \sys is three-fold:
(1) a \spad sidebar to facilitate exploration,
(2) cells movable between the notebook and the \spad to maintain organization,
and (3) linear execution with state forks to clarify program state.
An exploratory study (N=13) of open-ended data analysis tasks shows that 
\new{\sys features holistically promote clarity throughout a notebook's lifecycle, support realistic notebook tasks, and enable novel strategies for notebook clarity.}
\old
These results suggest that \sys supports maintaining clarity throughout the entirety of notebook authoring.
\end{abstract}




\begin{teaserfigure}
  \includegraphics[width=\textwidth]{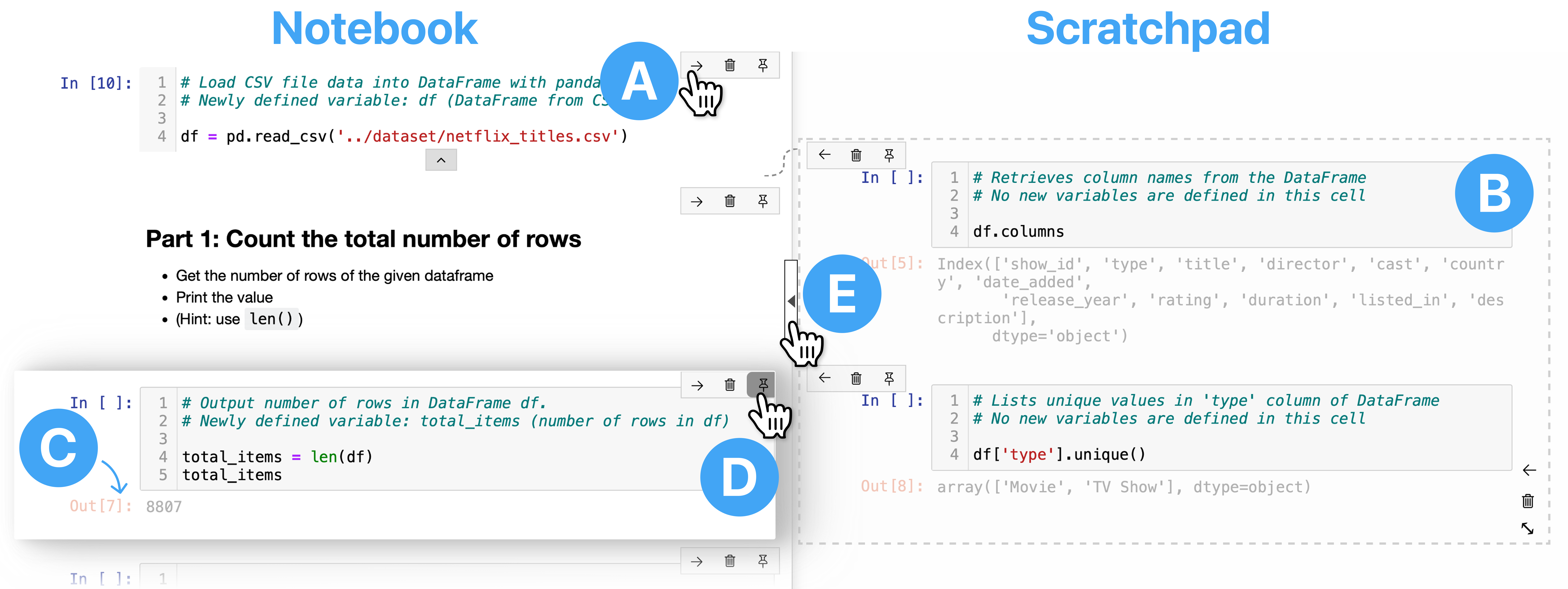}
  \caption{\sys divides a Jupyter notebook into clear content (left) and a \spad for exploration and references (right). \circledletter{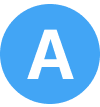}~A notebook cell can be moved to the \circledletter{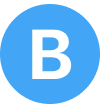}~\spad,
  where cell execution has no effect on the main notebook. 
  Linear execution ensures clarity in state, so rerunning a cell (\circledletter{A}) causes all subsequent cell outputs in the notebook and the \spad to gray out~\circledletter{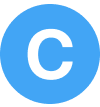}, indicating their staleness.
  \circledletter{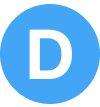}~One can pin a cell so that it floats on top of the screen during scrolling. 
  \circledletter{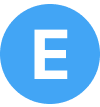} A toggle shows/hides the \spad as needed.
  Together, these features enable always-clear notebook authoring.
  }
  \label{fig:teaser}
  \Description{A screenshot of the Tidynote interface, including two side-by-side panels: "Notebook" on the left and "Scratchpad" on the right. The "Notebook" panel shows structured Python code and markdown for a data analysis task. A cell labeled 'A' contains code to load a Netflix dataset: df = pd.read\_csv('../dataset/netflix\_titles.csv'). A markdown cell describes "Part 1: Count the total number of rows" and detailed instructions. A cell labeled 'C' calculates and outputs the number of rows in df as 8807, with the output grayed out. Label 'C' has an arrow pointing at the output. The "Scratchpad" panel shows exploratory code cells. A cell labeled 'B' retrieves the column names from the DataFrame, with its output grayed out. The next cell lists the unique values in the 'type' column, showing 'Movie' and 'TV Show', also with the output grayed out. The labels with letters and hand cursors further highlight the interface's features: 'A' shows a hand cursor clicking on a rightward arrow on the upper right corner of the cell with the same label, which can move the cell from the Notebook to the Scratchpad. 'D' show a hand cursor clicking on a pin icon on the upper right corner of the cell with the same label, which can pin the cell so it floats on top of the screen. 'E' shows a hand cursor clicking on a toggle between the panels, which shows/hides the Scratchpad.}
\end{teaserfigure}

\maketitle

\section{Introduction}

% \todo{Authors of articles reporting on research involving human subjects or animals, including but extending beyond medical research, shall include a statement in the article that the research was performed under the oversight of an institutional review board or equivalent local/regional body, including the official name of the IRB/ethics committee, or include an explanation as to why such a review was not conducted. For research involving human subjects, authors shall also report that consent from the human subjects in the research was obtained or explain why consent was not obtained.}

% background of computational notebooks
% Literate programming~\cite{knuth1984literate} allows storytelling through code by interleaving code with explanatory text and other forms of media~\cite{granger2021jupyter}.
Computational notebooks, such as Jupyter~\cite{Jupyter}, have become widely popular among end-users by enabling storytelling through code~\cite{granger2021jupyter}.
Notebooks divide a program into \emph{cells} that can be interspersed with explanatory text, images, and formulas.
% Specifically, a notebook divides a program into \emph{cells} that can be interspersed with markdown annotations, images, and formulas.
Users can execute code cells individually, often out of order~\cite{lau2020design}, and thus easily perform exploratory programming~\cite{Kery2017:Exploring}.
Explorations are central to construct narratives in notebooks~\cite{keryStoryNotebookExploratory2018}, yet they are mostly small code cells~\cite{keryStoryNotebookExploratory2018, huang2025howscientists} and ``dirty tricks''~\cite{Rule2018:Exploration} that can quickly generate clutter, often with outdated and out-of-order results from prior runs.
A notebook filled with such explorations becomes \emph{messy}, difficult for one to understand its narratives~\cite{keryStoryNotebookExploratory2018} and in need of ``cleaning''~\cite{Rule2018:Exploration} to restore its clarity.
%\red{\emph{clarity}---structure, readability, and cleanliness.}

% limited support for clean explorations
A well-known challenge in notebooks is the tension between exploration and 
the need for
clarity~\cite{Rule2018:Exploration, liuRefactoringComputationalNotebooks2023a, Raghunandan2023:Code, keryStoryNotebookExploratory2018, huang2025howscientists}. Clarity is required in many phases throughout the notebook lifecycle,
% Notebooks are indeed used for many purposes that require clarity,
such as sharing results~\cite{subramanianCasualNotebooksRigid2020, huang2025howscientists}, tracking longitudinal findings~\cite{huang2025howscientists}, and providing tutorials~\cite{liuRefactoringComputationalNotebooks2023a}.
To ready a notebook for these purposes, prior systems support \emph{post-hoc} cleaning by turning portions of a notebook into slides~\cite{zheng2022tellingstories,li2023notable, wang2023slide4n}, cartoons~\cite{kang2021toonnote}, dashboards~\cite{Wang2022:StickyLand}, and videos~\cite{ouyang2024noteplayer}; collapsing cells into sections~\cite{ruleAidingCollaborativeReuse2018}; and gathering cells into complete slices~\cite{Head2019:Managing}.
However, these systems reinstate notebook clarity \emph{long after} the messes have been created.

Although some users may be content with post-hoc cleaning,
a recent study~\cite{huang2025howscientists} found that many users maintain clarity \emph{continuously}, while they edit a notebook, not just after.
They want to never let the mess get out of control.
However, neither Jupyter nor the post-hoc cleaning systems above provide direct support for continuous cleaning.
Consequently, these users adopt tedious manual workarounds, such as copying code to a fresh notebook for debugging, to maintain clarity during authoring~\cite{huang2025howscientists}.

% Moreover, making a notebook clear is often not the end of the notebook lifecycle---users iterate between explorations and activities that require clarity~\cite{huang2025howscientists}, relying on tactics such as informal version control~\cite{kery2017variolite} to develop new ideas while keeping old ideas intact.

% To support \emph{clarity}---organization, readability, and cleanliness---for, \eg, presentation from messy notebooks, prior systems turn portions of a notebook into slides~\cite{zheng2022tellingstories,li2023notable, wang2023slide4n}, cartoons~\cite{kang2021toonnote}, dashboards~\cite{Wang2022:StickyLand}, and videos~\cite{ouyang2024noteplayer}, collapse cells into sections~\cite{ruleAidingCollaborativeReuse2018}, and gather cells into complete slices~\cite{Head2019:Managing}.
% These systems make notebooks ready for presentation \emph{long after} the messes have been created.
% However, prior work~\cite{huang2025howscientists} shows that users want solutions and support for clarity while they are performing their explorations.
% % In fact, there are users who value clarity in their notebooks~\cite{huang2025howscientists}, yet this need has not been well supported.
% % does PageBreak count as part of these systems?
% % What, then, could be a better way for users who value notebook clarity to manage messes when authoring notebooks?

% In this paper, we explore a new approach that lets messes coexist with portions of the notebook that are always clear and presentable.

No prior systems support the needs of these users.
%They are helpless. They need to be saved. We will save them. Post hoc cleaning will not.
Post-hoc cleaning systems fail to confine and control messes \emph{as} they are created.
Therefore, we propose \emph{always-clear \del{computational notebooks}\new {notebook authoring}\old},
\del{a notebook augmentation}
\new{an authoring strategy }\old
that supports both clarity and exploration throughout the \emph{entire} notebook lifecycle,
implemented in a prototype named \sys for Jupyter notebooks~\cite{granger2021jupyter}. 
% \red{enable} the clarity of notebook content while maintaining support for exploration.
% The key idea behind \sys is to control and confine the messes \emph{as} they are created, not just after. 
In \sys,
continuous clarity is enabled via three mechanisms.
First, while the notebook captures the main narrative, a collapsible \emph{\spad} on the side stores work in progress, references, and one-off explorations (\autoref{fig:teaser}~\circledletter{B}).
% The \spad is divided into \emph{sections} that attach to portions of the notebook; this way, \spad cells can be grouped accordingly and confined to the main notebook from which they branch off.
%This space provides a content organization mechanism for maintaining clarity.
Second, cells can \emph{move between} the \spad and notebook as explorations progress into clear work or vice versa (\autoref{fig:teaser}~\circledletter{A}), with optional cell pinning to reduce scrolling (\autoref{fig:teaser}~\circledletter{D}).
The ability to move cells back and forth encourages fluid authoring without ``premature commitment''~\cite{CognitiveDimensions} to either clarity or exploration.
Third, cells are executed linearly, top-to-bottom, to align with the flow of information, with cells out of sync with the state grayed out (\autoref{fig:teaser}~\circledletter{C}).
This promotes clarity in program state.
%Additionally, when the user wants to refer to a cell while scrolling (the top activity in notebook work~\cite{huang2025howscientists}) in the notebook or the \spad, they could pin it so it floats on top of the screen (\autoref{fig:teaser}~\circledletter{D}).
%As an added convenience, \sys also offers cell pinning to reduce scrolling (\autoref{fig:teaser}~\circledletter{D}).
% This avoids repeated cells, such as the inspection of expressions.
Using these features, the main notebook can always be clear, in both structure and state: one can easily revisit their work from top to bottom or present their work to others without showing the messy explorations in the \spad.
Finally, \sys and standard Jupyter are mutually compatible: ordinary Jupyter functionality remains available in \sys, and notebooks created with \sys can be used in standard Jupyter.

We demonstrate how \sys supports always-clear notebook authoring via an exploratory study.
13 participants performed open-ended data analysis tasks using \sys with the goal of clearly encapsulating their work process for future reference.
Our findings show that all participants used all \sys features to perform their desired explorations \new{in realistic notebook tasks} \old while maintaining notebook clarity, \new{with various strategies for clarity enabled by \sys. }\del{ and rated it as a highly usable, low-effort system.}\old 
Even those participants who were previously careless about clarity were motivated to adopt clarity-driven workflows.
% Key findings include that participants leveraged all \sys features, adopted novel clarity strategies, and reported high confidence in clarity with low effort.\todo{more results}
% \red{participants found \sys to be usable, felt confident about and more aware of notebook clarity with relatively low effort in cleaning, and adopted multiple \sys-supported strategies for ensuring clarity.}
We conclude the paper with a discussion of notebook tasks that are best suited by always-clear notebooks and enumerate design opportunities for\del{further improvements} \new{future notebook systems and information systems}.\old

This paper contributes: 
\begin{itemize}
    \item \new \sys, a prototype system implemented as a Jupyter Notebook extension that supports the always-clear notebook authoring workflow;\old 
    % A proposal for always-clear computational notebooks that support both clarity and exploration throughout the entire notebook lifecycle, implemented in a prototype named \sys;
    \item A user study demonstrating \sys's effectiveness in supporting always-clear notebook authoring.
\end{itemize}

\section{Related Work}\label{sec:related}

Our goal is to simultaneously support clarity and nonlinear exploration in notebook authoring.
To this end, we review (1) challenges in using notebooks---particularly the tension between exploration and clarity; (2) systems for cleaning notebooks; and (3) alternative interfaces that support nonlinear exploration.

\subsection{Pain Points in Using Notebooks}\label{subsec:related-pain-points}
% user needs and challenges?

Computational notebooks such as Jupyter~\cite{granger2021jupyter} have become popular for data work, scientific computing, and machine learning~\cite{lau2020design, granger2021jupyter} due to their support for exploratory programming~\cite{Kery2017:Exploring} and narrating through code~\cite{knuth1984literate}.
Their flat, cell-based structure~\cite{huang2025howscientists} and nonlinear execution model~\cite{Perez2007:Ipython} not only distinguish them from traditional software development environments but also lead to unique usage challenges.

Challenges identified in prior work include low reproducibility~\cite{Pimentel2019:LargeScale, DeSantana2024:Bug} and poor code quality~\cite{grotov2022largeScale}. 
While these findings have motivated best practice recommendations similar to software development guidelines~\cite{Pimentel2019:LargeScale}, these recommendations often do not align with how notebook users actually work. 
For instance, Liu et al.~\cite{liuRefactoringComputationalNotebooks2023a} found that refactoring in notebooks differs largely from traditional software refactoring, and that notebooks with different focuses (\eg, exploration vs. exposition) exhibit different refactoring patterns.

Additional challenges concern user pain points, such as difficulty understanding program state, managing version history, and deploying to production~\cite{Chattopadhyay2020:Whats, subramanianCasualNotebooksRigid2020, huang2025howscientists, kery2017variolite}. 
These results suggest that notebook users work with different goals and requirements than those supported by software development tools, and yet their needs remain under-supported.

One possible explanation for the above challenges is the tension between exploration and clarity. 
Corpus studies~\cite{Rule2018:Exploration, liuRefactoringComputationalNotebooks2023a, Raghunandan2023:Code} and direct studies with users~\cite{Rule2018:Exploration, keryStoryNotebookExploratory2018, huang2025howscientists} highlight how notebooks range from exploratory work full of ``messes'' to artifacts with clear narratives, but the transition between these modes is effortful and often irreversible. 
Indeed, cleaning up exploratory work for clarity and presentation can require significant manual effort in content restructuring and code deletion, with users frequently relying on informal tactics like commenting out code, creating fresh notebooks, or sectioning via markdown headings~\cite{kery2017variolite, keryStoryNotebookExploratory2018, huang2025howscientists}. 
This tension is exacerbated in collaborative settings, where disorganized notebooks hinder shared understanding~\cite{wangHowDataScientists2019}. 
In reality, a notebook's lifecycle can be filled with back-and-forth iterations between exploratory work and intermediate presentations, and users have created many strategies that adapt to their needs of clarity: some users delay cleanup until sharing~\cite{Rule2018:Exploration}, some always prioritize finding solutions over maintaining code quality~\cite{kery2017variolite}, while others always keep a notebook clear even when not collaborating or creating artifacts~\cite{huang2025howscientists}.
Our work aims to embrace the spectrum of notebook clarity by simultaneously supporting clarity and exploration throughout the entire notebook lifecyle.

\subsection{Systems for Cleaning Notebooks}\label{subsec:related-systems-for-cleaning}

Some systems aim to help the user create cleanliness in their notebook. Some provide extra structure to explain notebook organization and some help distill a notebook into a streamlined form for presentation. These are best suited for \emph{post hoc} cleaning. We discuss these systems below, as well as StickyLand~\cite{Wang2022:StickyLand}, which, like our envisioned system, supports cleaning \emph{during} notebook authoring rather than just post hoc.

Several systems augment the notebook structure to facilitate clear organization
of work~\cite{ruleAidingCollaborativeReuse2018, chattopadhyay2023makemakesenseunderstanding, lin2025InterLink}.
These structure-focused systems allow users to collapse or hide groups of cells under named headers~\cite{ruleAidingCollaborativeReuse2018}, annotate sections~\cite{chattopadhyay2023makemakesenseunderstanding}, or link explanatory text to corresponding code and outputs~\cite{lin2025InterLink}.
However, these systems provide an overhead in establishing notebook structure in exploratory settings; more importantly, these techniques are best suited for \emph{post hoc} organization because the overhead needs to be repeated as a notebook evolves.
Our work aims to maintain clarity in notebook organization \emph{during} fluid exploration with minimal overhead.

Another class of tools turn notebooks into presentable formats.
From notebooks, prior systems auto-generate slide decks~\cite{zheng2022tellingstories} and associated bullet points~\cite{wang2023slide4n}, dashboards with storytelling annotations~\cite{li2023notable}, data comics~\cite{kang2021toonnote}, or videos~\cite{ouyang2024noteplayer}, but they assume the notebooks are already clear and organized.
Code gathering tools~\cite{Head2019:Managing} slice execution logs to produce a minimal, presentable notebook slice from a selected output; while this reduces clutter within one notebook, repetitive cells could persist across notebook slices, and cleaned slices quickly fall out of sync with ongoing edits in the original notebook.
Compared to the above systems, which focus on \emph{post hoc} mess management to support clear presentations from messy explorations, we propose controlling notebook mess \emph{during} exploration while enabling easy transition to presentation.

Perhaps closest to our work in terms of functionality, StickyLand~\cite{Wang2022:StickyLand} allows pinning cells in a floating dashboard on top of a messy notebook, maintains provenance between the source notebook and the dashboard, and enables quick dashboard-based presentation without explicit cleaning.
However, StickyLand differs from our work in design philosophy: ``sticky'' cells are flexible, but ambiguous in their purpose because they can range from transient reference material to presentable results, whereas \sys's \spad is less ambiguously intended for scratch work. If a StickyLand user wants the main notebook to be their presentable artifact instead of the sticky space, then the sticky space is less useful because it does not make sense to put all exploratory code into a floating space.
\del{In addition, StickyLand reuses standard Jupyter execution semantics and was thus not designed to promote clarity in program state, which is a design goal of \sys.}%
\old
We analyze StickyLand and other closely-related systems in more depth in \autoref{sec:design-goals}, in relation to our design goals.

\subsection{Notebook State Management and Exploratory Interfaces}\label{subsec:related-non-linear-exploration}

Some non-Jupyter systems provide alternative state management models to mitigate state messes~\cite{Polynote, Observable, marimo, rawn2025Pagebreaks}. Providing automatically managed state, such as automatically rerunning dependent cells, can alleviate state confusion, but does not necessarily promote exploration or provide a way to separate presentation-ready content from archival or temporary work.
Thus, other systems aim to improve nonlinear exploration itself.
For traditional scripting, Variolite puts in situ version control on code blocks to facilitate variants and comparisons~\cite{kery2017variolite}, but does not have the benefit of cell-based execution as in notebooks.
In the context of notebooks, Fork~It~\cite{weinmanForkItSupporting2021} enables state forking and backtracking and displays branches side by side, and Kishuboard~\cite{fang2025Kishuboard} brings version control to the data level.
However, the clarity of program state can still remain problematic in these systems due to the standard nonlinear execution.
2D canvas nodes-and-wires notebook systems like natto~\cite{natto} let users spread alternative explorations along a second spacial dimension, but wires can become messy ``spaghetti''~\cite{VisualProgrammingTheOutlookFromAcademiaAndIndustry}, and organizing the 2D space requires deliberate and careful user intervention.
Instead of building another 2D nodes-and-wires system that could lead to spaghetti, we are
inspired by the inline annotation interface of TextTearing~\cite{yoon2013texttearing} and in-situ branching interfaces~\cite{sungwon2024evaluating, weinmanForkItSupporting2021, kery2017variolite}
to preserve the original linear display of cells as much as possible. We thus minimally augment the linear cell order with \emph{one level of branching} to allow for intuitive nonlinear exploration.

\section{Always-Clear \del{Computational Notebooks}\new{Notebook Authoring}\old}\label{sec:design-goals}

\begin{figure*}[t]
  \centering
  \includegraphics[width=\textwidth]{images/table-revised.png}
  \caption{\new
  \emph{Comparative analysis of different strategies for managing notebook messes}.
  We compare the mechanisms of three existing strategies---out-of-notebook cells~\protect\cite{spadextension, Wang2022:StickyLand}, post-hoc cleaning~\protect\cite{ruleAidingCollaborativeReuse2018, Head2019:Managing}, and state forks~\protect\cite{weinmanForkItSupporting2021}---with our envisioned \emph{always-clear notebook authoring}, across five actions relevant to keeping a notebook clear throughout its lifecycle---performing exploration (global and cell-based), clearing unused exploration, iterating between exploration and clarity, and reducing state messes. 
  ``No additional support'' means the actions are achievable manually, but the corresponding approach does not provide additional support beyond the base, traditional Jupyter; ``Disallowed'' means such actions are unattainable at all.
  Our approach is featured in the last row, the only strategy that supports keeping a notebook clear throughout its entire lifecycle.\old
}
  \label{fig:big-table}
  
  \Description{
  This is a table with six columns and six  rows. 
  Columns are types of User Action, relevant sub-actions, and examples.
  Column 1 and 2: Performing Exploration, with sub-actions "global" in Column 1 (example: "An irrelevant exploration") and "cell-based" in Column 2 (example: "An exploration based on a cell").
  Column 3: Creating Clarity (example: "Preparing for presentation").
  Column 4: Juggling between Exploration and Clarity (example: "Manager with guests suddenly walks in wanting an update").
  Column 5: Reducing State Messes (example: "Restoring data after out-of-order execution).
  Column 6: Keeping Clarity Throughout (example: "Keeping a notebook always clear in content and state").  
  Rows are Strategies and relevant tool names.
  Row 1 and 2: Out-of-Notebook Cells, with tools "scratchpad extension, citation number 36" in Row 1 and "StickyLand, citation number 46" in Row 2.
  Row 3 and 4: Post-hoc Cleaning, with tools "Janus, citation number 37" in Row 3 and "Gather, citation number 14" in Row 4.
  Row 5: State Forks, with tool "Fork It, citation number 47".
  Row 6: Always-Clear Authoring, with tool Tidynote (italic, golden fonts, main contribution of this paper).
  Each table cell is a blue block that shows how the tool under a strategy supports a user action, or a dash indicating no support.
  Row 1 (scratchpad extension) has one blue block only under Column 1 - "Performing Exploration - global": A hidable cell on the side.
  Row 2 (StickyLand) has two blue blocks under Column 3 - "Creating Clarity": A floating window, and Column 4 - "Juggling between Exploration and Clarity": Move between NB \& floating window.
  Row 3 (Janus) has one blue block only under Column 3 - "Creating Clarity": Hide cells to the side.
  Row 4 (Gather) has one blue block only under Column 3 - "Creating Clarity": Slice cells to new NB.
  Row 5 (Fork It) has two blue blocks under Column 1 - "Performing Exploration - global": Create cell branches, and Column 3 - "Creating Clarity": Delete branches.
  Row 6 (Tidynote) has blue blocks under all columns: "SPad on side" for "Performing Exploration - global", "Cell-based scratch work" for "Performing Exploration - cell-based", "Hide SPad" for "Creating Clarity", "Move cells between NB \& SPad" for "Juggling between Exploration and Clarity", "Linear execution \& state forks" for "Reducing State Messes", and "SPad, movable cells, \& linear execution" for "Keeping Clarity Throughout".
  }
\end{figure*}

The goal of our work is to support the clarity of a notebook \emph{throughout its lifecycle}---from initial creation, to any number of intermediate presentations, and to continued editing after each presentation. 
This goal entails clarity requirements in both the content and the program state.

While there are approaches attempting to manage notebook clarity, primarily after messes are created, there are no prior systems that fully achieve the wholistic nature of the above goal.
\new
To inform the design of our envisioned ``always-clear'' notebook authoring, we first examine how existing tools and practices manage the tension between exploration and clarity during notebook authoring. 
We analyze three representative strategies---out-of-notebook cells, post-hoc cleaning, and state forks, examining how each supports and/or fails to support the clarity needs throughout the notebook authoring process (\autoref{subsec:comparative-analysis}). 
This analysis reveals three critical weaknesses that prevent existing approaches from keeping notebooks \emph{consistently} interpretable and presentable. 
These weaknesses directly motivate our design goals and the design decisions behind \sys  (\autoref{subsec:design-goals}).

\old

\subsection{Comparative Analysis}\label{subsec:comparative-analysis}

\new
Through the literature~\cite{Rule2018:Exploration, keryStoryNotebookExploratory2018, Chattopadhyay2020:Whats, huang2025howscientists}, we first identify five main user actions in keeping a notebook clear throughout its life cycle: structuring exploration (1) globally and (2) cell-wise, (3) cleaning up exploration, (4) maintaining exploration while keeping the notebook interpretable, and (5) reasoning about program state.
We review how each relevant prior strategy~\cite{spadextension, Wang2022:StickyLand, ruleAidingCollaborativeReuse2018, Head2019:Managing, weinmanForkItSupporting2021} supports these user actions or fails to do so.

To explain how our proposal compares to prior approaches, \autoref{fig:big-table} shows our comparative analysis, showing the above user actions as column headers and the reviewed strategies in each row; our approach is shown in the last row.
We now go into more details of each strategy and their support for each relevant user action.

\old

\mypara{Out-of-Notebook Cells}
\new
Both the Jupyter scratchpad extension~\cite{spadextension} and StickyLand~\cite{Wang2022:StickyLand} introduce auxiliary spaces---a side cell or floating windows---to give users room to explore without cluttering the main notebook. While these structures offer an external workspace, they do not improve the clarity of the main notebook itself, which remains the primary ``battlefield” where users read, write, and present results.
Furthermore, these systems address content clarity but not state clarity. They retain Jupyter’s global execution semantics: all cells, regardless of their location, share the same underlying state. Exploratory code in a scratchpad or floating window can therefore modify variables and outputs throughout the notebook, disallowing any exploration that is meant for just individual cells.
As such, users remain responsible for preventing state inconsistencies with these approaches, limiting the usefulness of these auxiliary spaces for maintaining overall clarity.

\old

\mypara{Post-hoc Cleaning}
While providing no additional support for exploration, post-hoc cleaning approaches allow creating clarity \emph{after} exploration: Janus lets users move cells to a folded tab on the side~\cite{ruleAidingCollaborativeReuse2018}, while Gather allows slicing out a subset of cells into a separate notebook~\cite{Head2019:Managing}.
These systems envision cleanup as a final step, rather than an intermittent process, so their support for juggling between exploration and clarity is limited to manual copy/paste of code: Janus disallows moving the folded side-cells back, and slices out of Gather are independent from the source notebook.
The post-hoc nature of the cleanup does not suit users who want to maintain clarity \emph{throughout} authoring~\cite{huang2025howscientists}.

\mypara{State Forks}
Fork~It~\cite{weinmanForkItSupporting2021} supports state branching, where each branch contains a fork of the notebook state.
Users compare branches during exploration and remove unnecessary branches to restore clarity.
However, only one branch point is allowed at a time, so explorations must be deleted before creating branches at a different point. This limitation discourages juggling between exploration and clarity, and impedes the user from archiving explorations.
And although each branch forks the notebook state at the branch point, Fork~It keeps the default Jupyter execution semantics where one can run cells out of order, which led to confusion in their user study.
Because of these limitations in branching and state clarity, maintaining clarity throughout notebook authoring still requires considerable manual effort.

\mypara{\new{Summary of Insights \& Weaknesses}}
\new{Taken together, these strategies provide valuable insights into promoting clarity in notebook---auxiliary exploration spaces, post-hoc cleaning mechanisms, and state forking. However, we also }
\old
identify three key weaknesses that keep these prior techniques from enabling clarity throughout a notebook's \emph{entire} lifecycle.

\spara{Lack of flexible, structured exploratory spaces} 
In practice, exploratory work is often scoped to certain cell(s) (which is part of the motivation for PageBreaks~\cite{rawn2025Pagebreaks}).
Existing auxiliary spaces operate only at the global level and remain decoupled from the cells where exploration originates, both visually and state-wise, offering no structured way to support cell-level or multi-scale explorations.

\old

\spara{\new{Insufficient support for fluid iteration between exploratory and clear work}}
\old
During notebook authoring, exploration might mature into clear content, while previously clear content could regress to exploration (\eg, archived but not removed~\cite{keryStoryNotebookExploratory2018, huang2025howscientists}).
Such limited support thus contradicts the dynamic nature of how a notebook evolves and inhibits always-clear authoring.

\spara{No support for ensuring clarity in program state}
All existing strategies reuse the default cell execution mechanism in Jupyter notebooks that could result in inconsistent state and unreproducible code~\cite{Pimentel2019:LargeScale, wangHowDataScientists2019}.
In the long term, these challenges
lead to difficulty in understanding state and in longitudinal notebook work.

\subsection{Design Goals}\label{subsec:design-goals}

\new
Based on the weaknesses above, we derived the following design goals for always-clear notebook authoring, and we explain how each goal informed the corresponding \sys design decisions.
\old

\mypara{\DGOne}
Exploratory work should be easy to \new initiate, localize, and \old organize via temporary and flexible structures, specific to a cell or general to the entire notebook, to facilitate its progression into clear work. 

\new
\sys introduces a \spad attached to the notebook, where each exploratory section is initialized by moving a specific cell into the scratchpad. Cells can also be pinned to remain visible while scrolling, supporting multi-step reasoning without losing context.  These structures let users keep relevant explorations close to their originating cells.
\old

\mypara{\DGTwo}
Cells should be easy to move between exploratory and clear portions of a notebook as it matures.

\new
 \sys enables bidirectional movement of cells between the main notebook and the \spad.
 This supports the natural progression and flow of notebook work, allowing explorations to be hidden or promoted without tedious manual effort.
\old

\mypara{\DGThree}
\new
Execution should behave predictably and align with the notebook’s visual layout, preventing hidden or inconsistent states.

\sys implements linear execution semantics and state forking aligned with the user’s visual structure of notebook work. This maintains reproducible state while still allowing exploratory divergence when needed.
\old

\smallskip\noindent
\new
Guided by these decisions, we implement \sys as an extension to Jupyter notebook.
\old
\autoref{sec:walkthrough} demonstrates the use of \sys, and \autoref{sec:implementation} details its technical implementation.

\section{\sys Walkthrough}\label{sec:walkthrough}

Ali is a data scientist who also teaches introductory data analysis with Jupyter notebooks at a university. 
Today, she is developing a notebook that analyzes a dataset of Netflix shows as lecture material.\footnote{This is one of the tasks from our user study, and Ali's story and interactions with \sys are based on the experience of some participants}
As this notebook is meant to be
shared with students, Ali wants it to be clear while also 
exploring different ways of designing the material.
As such, she decides to use \sys, 
an always-clear notebook system for Jupyter.\footnote{Demo video in supplementary and \url{https://tinyurl.com/tidynote-study-rep}}

\begin{figure*}[t]
  \centering
  \includegraphics[width=\linewidth]{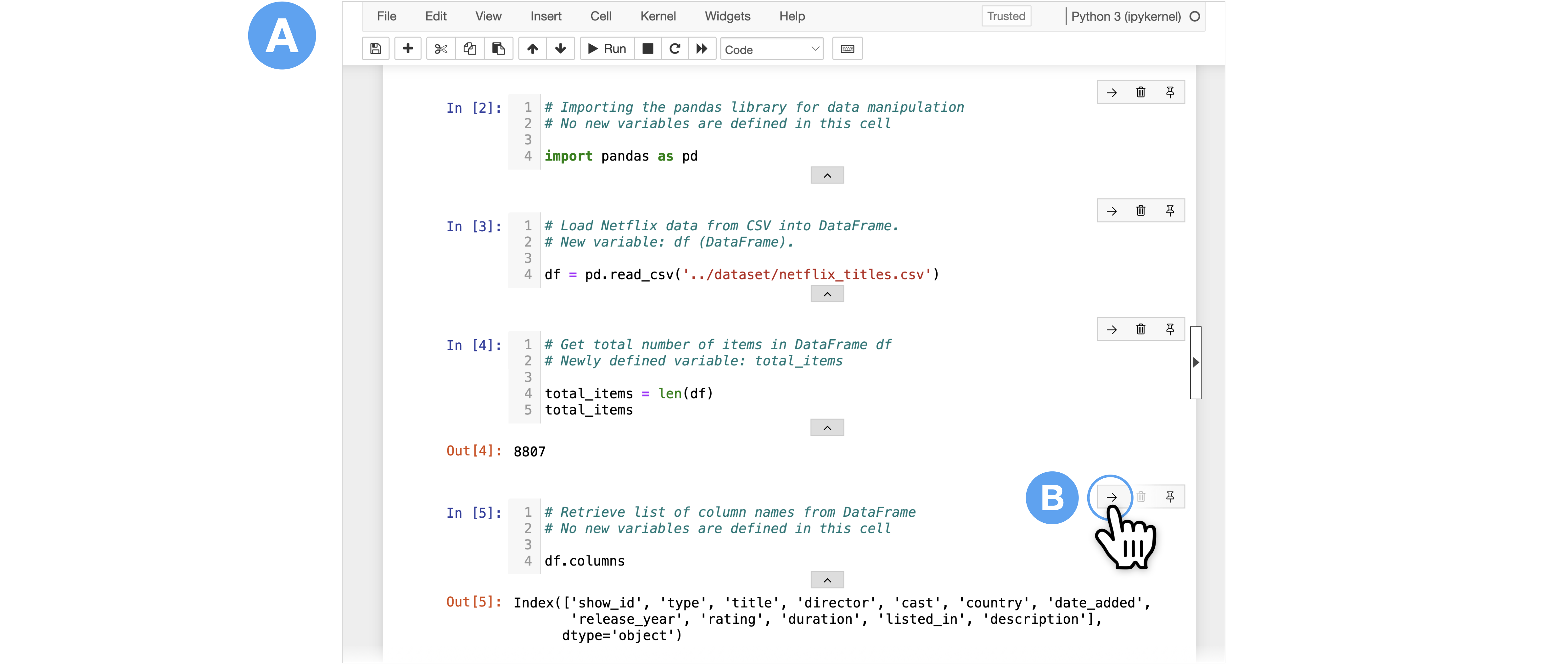}\\[1em]
  \includegraphics[width=\linewidth]{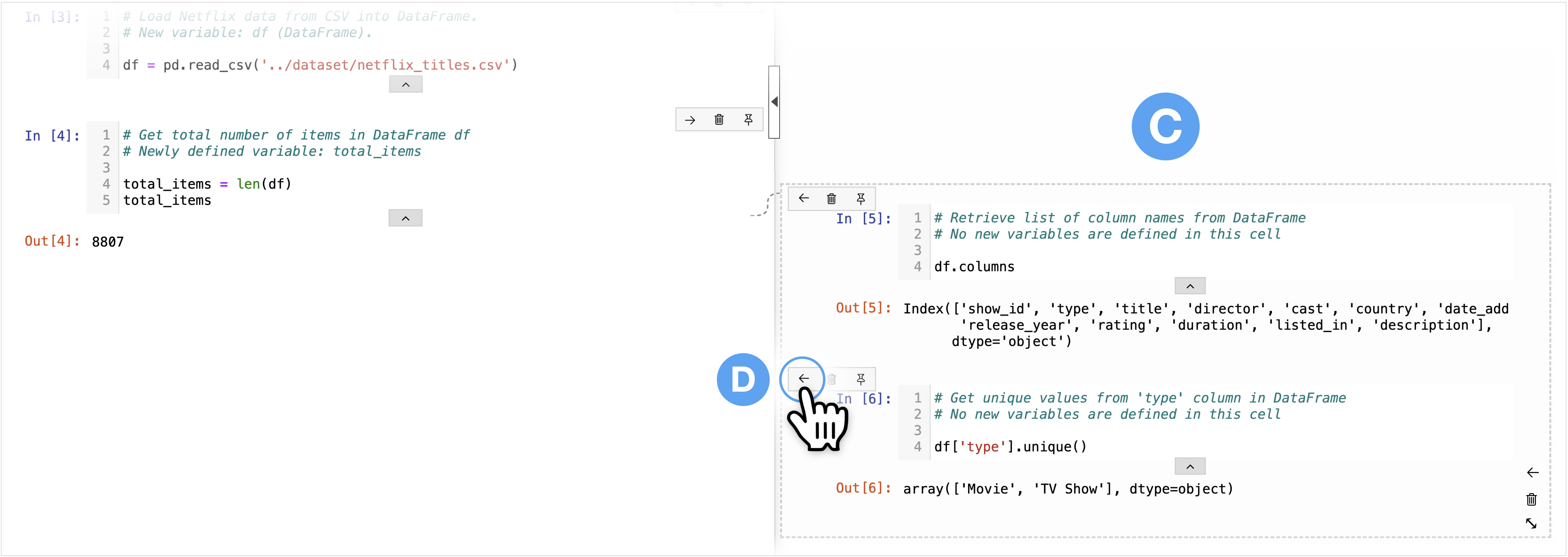}
  \caption[]{
    \circledletter{A} The full \sys interface, showing Ali's initial notebook with the \spad hidden.
    \circledletter{B} Clicking the \raisebox{-0.23\height}{\includegraphics[height=1em]{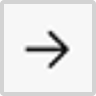}} button expands the right margin to reveal the \circledletter{C} \spad and moves the cell into it.
    \circledletter{D} The \raisebox{-0.23\height}{\includegraphics[height=1em]{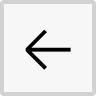}} button moves a cell back to the notebook.
    (Each~code cell starts with two lines of automatically generated comments.)
    }
  \label{fig:overview-interface}
  \Description{
    (A) A standard Jupyter notebook with four code cells. The upper right of each cell has three buttons: a right arrow, a trash can, and a pin.
    (B) A pointing cursor is over the right arrow button of the fourth last cell (the cell with `df.columns`).
    Second subfigure follows below. The right margin is now open. The main notebook no longer contains the `df.columns` cell.
    (C) In the right margin, two cells are surrounded by a dotted border (a scratch section with two cells). The first cell is `df.columns`. The second is `df['type'].unique()`. A short dotted line connects the dotted border to under the third cell of the main notebook, showing where the scratch section is anchored. The triplet of buttons on each cell are now on the upper left, with a left arrow instead of the right arrow.
    (D) A pointing cursor over the left arrow button of the `df['type'].unique()` scratchpad cell, about to click to move the cell back to the main notebook.
    }
\end{figure*}

\begin{figure*}[t]
  \centering
  \includegraphics[width=\textwidth]{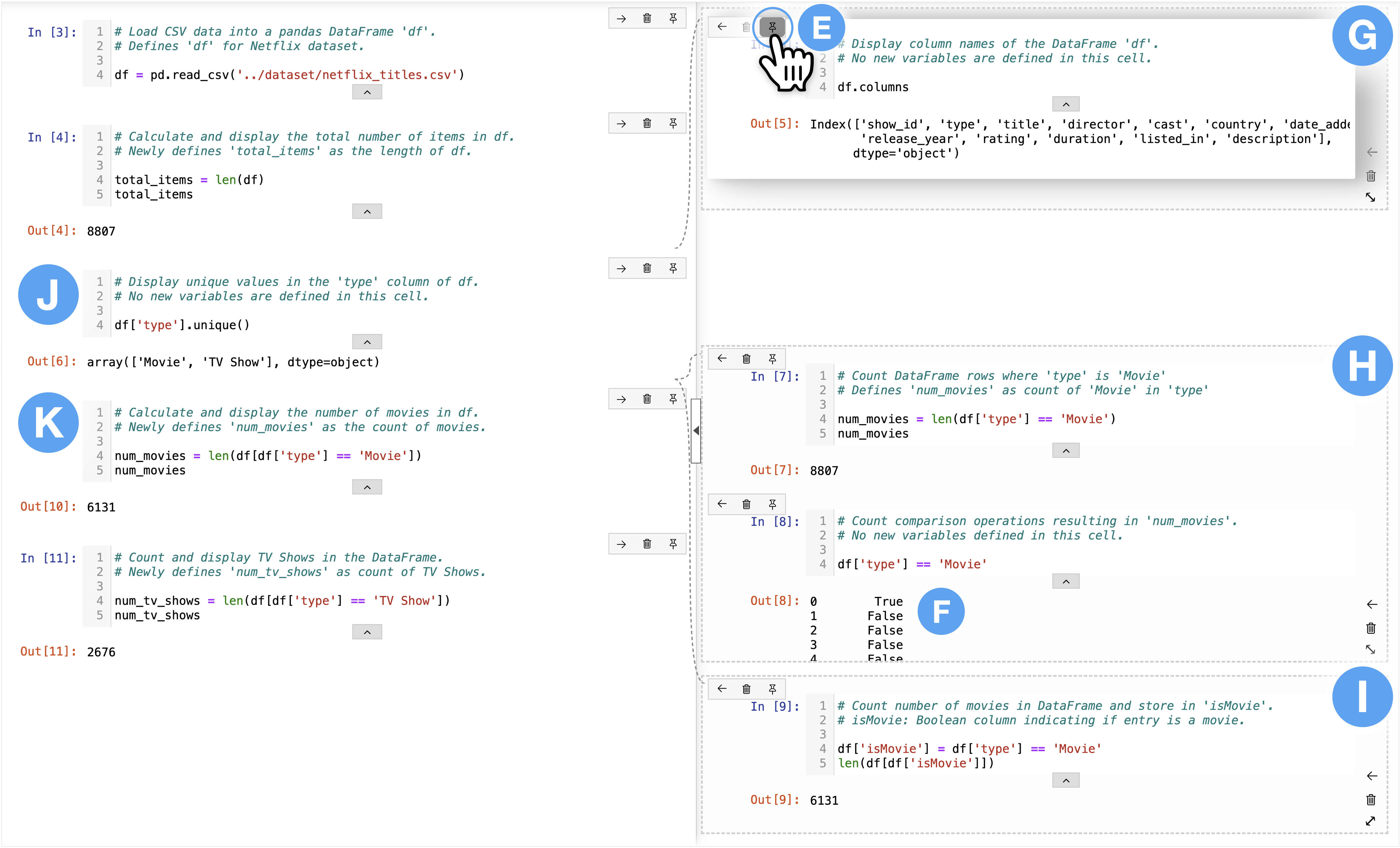}
  \caption{Cells can be pinned~\circledletter{E} to prevent them from scrolling offscreen. The \spad can contain multiple \mbox{scratch sections~\circledletter{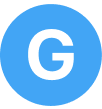}~\circledletter{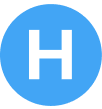}~\circledletter{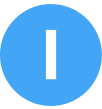}}, each with independent state from each other and from the main notebook. Multiple scratch sections might branch from the same main notebook cell, \eg \circledletter{H}~and~\circledletter{I} from~\circledletter{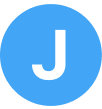}.
  }
  \label{fig:overview2}
  \Description{
  The notebook and scratchpad from the prior figure has more cells now, with several lettered callouts.
  (E) A pointer over the pin button on the `df.columns` cell. The cell is now floating and pinned to the top of the screen in the scratchpad. This single cell is in its own scratch section labeled (G).
  The next scratch section labeled (H) has two cells, the first with code `num_movies = len(df['type'] == 'Movie')` `num_movies` with an output of 8807, the same as `total_items`, the second cell has code `df['type'] == 'Movie'` outputting a boolean series labeled (F).
  The final scratch section labeled (I) contains one cell with the code `df['isMovie'] = df['type'] == 'Movie'` `len(df[df['isMovie']])` and an output of 6131.
  The (H) and (I) scratch sections are connected by curved dotted lines to just below the `df['type'].unique()` cell in the main notebook, labeled (J), indicating where their state branches from.
  The cell after (J), labeled (K), has code `num_movies = len(df[df['type'] == 'Movie'])` `num_movies` with output 6131.
  A final cell below it has code `num_tv_shows = len(df[df['type'] == 'TV Show'])` `num_tv_shows` with output 2676.
  }
\end{figure*}

\mypara{\Spad}
The display of \sys behaves like an ordinary Jupyter notebook, as shown in \autoref{fig:overview-interface}~\circledletter{A}. 
Ali loads her dataset into a Pandas dataframe and computes the number of items in the dataset with \scode{total_items = len(df)}.
From this starting point, Ali would like to teach her students how to count the number of rows matching a particular condition.
She needs to explore to find a suitable column and condition for the demonstration, so she creates a new cell and writes \scode{df.columns} to list the columns in the dataset.
She does not want this exploration to be part of the main narrative of this notebook. 
\sys provides a \emph{\spad} in the notebook's right margin: a parallel set of cells for holding temporary, old, exploratory, or alternative computations that the user does not want to include in the main narrative.
In the upper right corner of the \scode{df.columns} cell, Ali presses the \raisebox{-0.26\height}{\includegraphics[height=1.2em]{images/buttons/moveToSpad-b.png}} right arrow button~\circledletter{B} which expands the right margin of the notebook to reveal the \spad and moves the cell into it \circledletter{C}.
Cells in the \spad behave as normal, but are ``aside'' from the main notebook narrative.
A dotted line on the cell connects it to the main notebook.
A \spad cell can reference variables defined in the main notebook before this line, but new and changed variables in the \spad do not affect the main notebook.
The user may also hide the whole \spad, leaving only the main narrative visible.

Ali thinks the columns \scode{'type'} and \scode{'release_year'} might be useful. She clicks Jupyter's \raisebox{-0.2\height}{\includegraphics[height=1.2em]{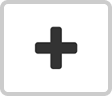}} in the top tool bar that adds a cell below the active cell, in this case, in the \spad below \scode{df.columns}, allowing her to continue her exploration outside the main narrative. She writes \scode{df['type'].unique()} to discover there are only \scode{'Movie'} and \scode{'TV Show'} types in this data set. She adds another cell to similarly check \scode{'release_year'} and sees many more unique values. Ali decides to go with the \scode{'type'} column. She deletes the cell checking the unique values of \scode{'release_year'}.

Ali will include \scode{df['type'].unique()} in her lecture, so \circledletter{D} she presses the \raisebox{-0.26\height}{\includegraphics[height=1.3em]{images/buttons/moveToNB-b.png}} button to move that cell from the \spad back to the main notebook.

\mypara{Cell Pinning}
She expects it may be useful to reference the column names later. To do so without needing to scroll back in the notebook, she presses the pin button \raisebox{-0.26\height}{\includegraphics[height=1.2em]{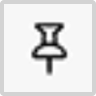}} on the \scode{df.columns} cell, which causes the cell to float and stick to the top or bottom of the screen when it would otherwise scroll off the viewport (\autoref{fig:overview2}~\circledletter{E}).

\mypara{Sandboxed Scratch Sections}
In a new cell in the main notebook, Ali writes \scode{num_movies = len(df['type'] == 'Movie')} hoping to count the number of rows of the movie type. However, upon inspecting the variable, she realizes that she made a mistake: \scode{num_movies} outputs the same value as \scode{total_items}. She is confused about the mistake and presses the  \raisebox{-0.26\height}{\includegraphics[height=1.2em]{images/buttons/moveToSpad-b.png}} button to pull out the cell to the \spad, expecting to potentially need several cells to debug.
% , where a scratch section can have several cells).
Testing the output of just \scode{df['type'] == 'Movie'} shows that her code did not filter out a subset of the data based on the condition, instead just producing a list of boolean flags (\autoref{fig:overview2}~\circledletter{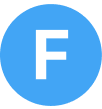}).
She thinks she might want to save these flags as a new column, but she does not (yet) want to mutate \scode{df} in the main notebook or for the existing scratch cells.
She creates a new cell in the notebook and moves it to the \spad: this creates a new ``scratch section'' whose state is independent of other sections, and changes to its state will not affect the main notebook. The scratchpad in \autoref{fig:overview2} has three scratch sections \circledletter{G}~\circledletter{H}~\circledletter{I}, each with its own gray dotted border.
Multiple scratch sections can branch off the same main notebook cell: as shown by the dotted lines, both \circledletter{H} and \circledletter{I} shared the same parent cell \circledletter{J}, but the state of \circledletter{H} and of \circledletter{I} are independent of each other.

In the new scratch section \circledletter{I} Ali adds an \scode{isMovie} column to \scode{df} with \scode{df['isMovie'] = df['type'] == 'Movie'}.
Ali then remembers she can count after filtering, which she tries with \scode{len(df[df['isMovie']])}.
This produces the desired result, but now she decides that creating a new column just to count may be a bit excessive. Happily, scratch sections are sandboxed so her mutated dataframe is scoped only to \circledletter{I}: \eg, re-running \scode{df.columns} in \circledletter{G} will not show the new column, nor will the abandoned exploration pollute the main notebook.

Ali settles on \scode{num_movies = len(df[df['type'] == 'Movie'])}. She adds this as a new cell~\circledletter{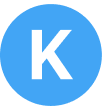} in the main notebook.
She repeats the computation with the \scode{'TV Show'} type of shows. She confirms, using the \spad, that the sum of the two values is indeed the same as \scode{total_items} (not shown).

\begin{figure*}[t]
  \centering
  \includegraphics[width=.65\linewidth]{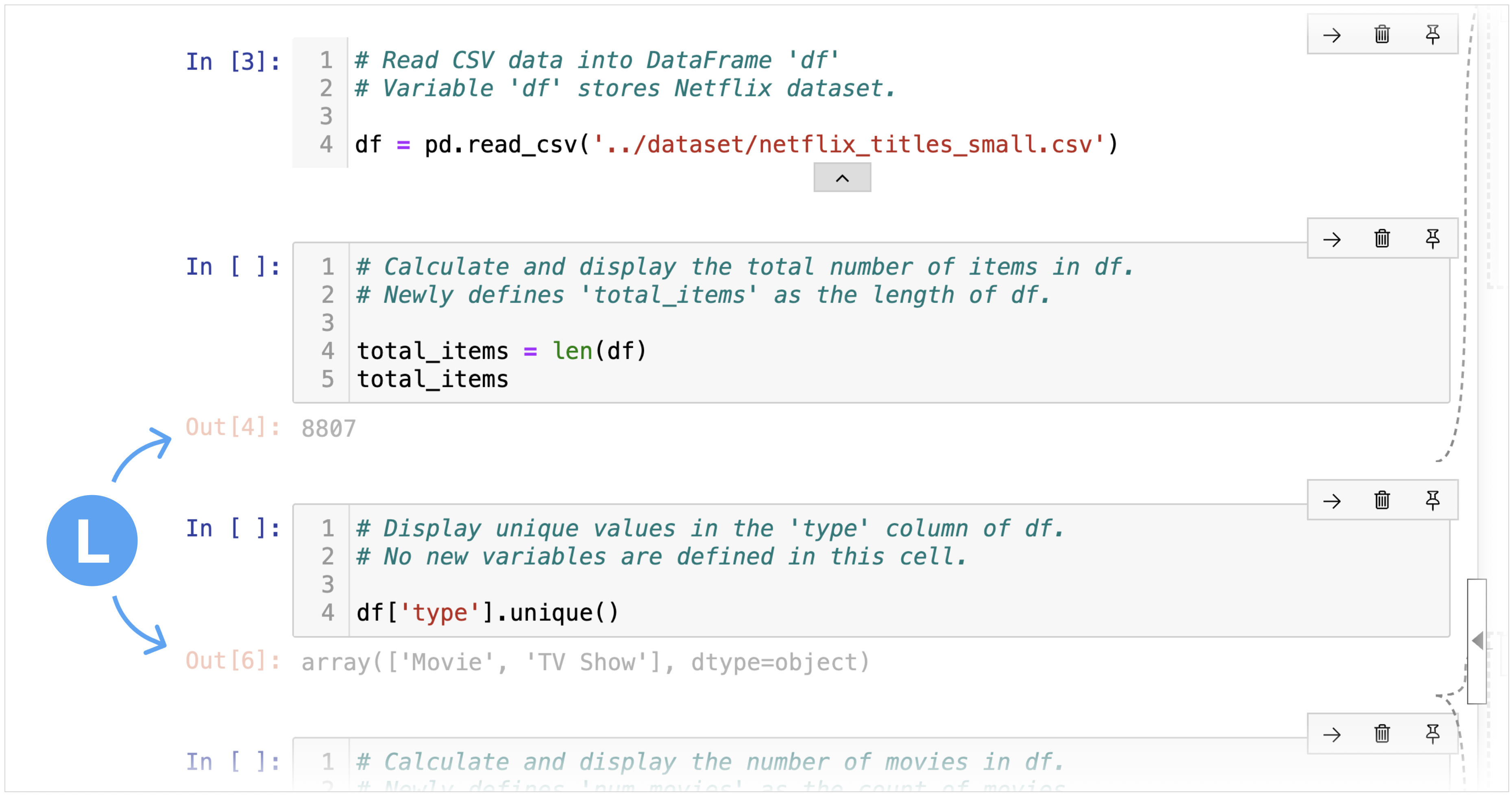}
  \caption{Rerunning the first cell (\scode{df = ...}) is a nonlinear execution: the output of later cells are grayed out \circledletter{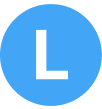} to indicate staleness.
  }
  \label{fig:results-cleared}
  \Description{
  The first three cells from the main notebook in the previous figure shown again.
  The first cell was re-executed, causing the outputs of the subsequent cells to gray out (L).
  }
\end{figure*}

\mypara{Linear Execution}
Ali decides the raw input data is too large to distribute to her students, so in a separate text editor she removes most of the CSV rows from the input file. In her notebook she re-runs the cell with \scode{read_csv}. \sys enforces linear execution to insure integrity of the notebook narrative: all subsequent cells in the main notebook and scratchpad have their outputs grayed out to indicate they are stale and must be re-run to update (\autoref{fig:results-cleared}~\circledletter{L}). Ali does so.

\mypara{Sharing}
Although most of Ali's cells are short, long cells are common.
Like JupyterLab notebooks, \sys supports cell folding, but instead of showing the default ``$\cdots$'' after folding, \sys collapses the cell to only its first two lines: the first line displays an AI-generated summary of the cell's operation and the second lists any new variables defined in the cell.\footnote{Implementation with GPT-4o~\protect\cite{GPT-4o} detailed in~\autoref{app:folding}}
Before sharing the notebook with her students, Ali presses collapse buttons \raisebox{-0.18\height}{\includegraphics[height=1em]{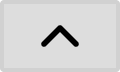}} to fold the cells that import Pandas and load the CSV, as the students are already familiar with this code (collapsing not shown in figure).
Finally, Ali saves the notebook as is and shares it. The students can open it in \sys and, by default, see only the notebook portion with Ali's explorations in the \spad hidden away,
but they can still open the \spad and experiment on their own for more practice.

\mypara{Summary} 
Ali developed a data analysis notebook for instruction using \sys to simultaneously explore ideas while keeping the main notebook portion for presentation clean.
The primary enabler of this dual-use was \sys's \emph{scratchpad}, a space that allows for side thoughts, explorations, alternative computations, and archived old ideas to separate from the notebook's main narrative. The scratchpad, along with \emph{cell pinning} for quick reference, enabled Ali to explore and experiment with a record of her work she could consult now (with pinning) or later, without muddying her public-facing narrative (\emph{\DGOne}).

With the ability to move cells between the notebook and the \spad, Ali could carefully craft that public-facing narrative within her notebook without requiring an explicit cleaning phase of her notebook authoring. She could move explorations or archived code to the \spad and hide the mess at any time to immediately present the content to her students or co-teachers (\emph{\DGTwo}).

By enforcing \emph{linear execution}, with each \emph{scratch section} branching off from the main narrative in a sandbox,  \sys eliminates the overhead of reasoning about out-of-order notebook state, freeing Ali to explore fearlessly, \eg, mutating the dataframe as she worked out the best way to count rows (\emph{\DGThree}).

Had Ali not used \sys, she would have spent more time deleting exploratory cells, more time scrolling to reference cells (instead of pinning them), more mental energy thinking about the kernel state, more time re-running prior cells to reset her dataframe after mutating it, and likely more time re-writing code she had previously deleted (because she could not stash it). \sys streamlines the process of keeping a tidy notebook.

\section{\sys Implementation}\label{sec:implementation}

\sys is an extension of Jupyter Notebooks v6 written in about 4,500 lines of TypeScript for GUI rendering and about 250 lines of Python for code execution.\footnote{\sys source and examples: \url{https://github.com/rlisahuang/tidynote-artifact}}
We implement cell pinning and folding with CSS manipulation.
Below we explain \spad rendering, linear execution, \newR{indicating stale cells, }\old compatibility with standard Jupyter, and generalizability.

\mypara{\Spad Rendering}
The \spad is placed next to the main notebook and contains scratch sections attached to the main notebook, similar to text annotated with comments in a rich text editor.
A scratch section must be initialized with a notebook cell moving into the \spad: given two consecutive notebook cells $C_k$ and $C_{k+1}$, when $C_{k+1}$ is moved to the \spad, it is placed in a new scratch section attached to $C_k$. 
All sections attached to $C_k$ will be displayed in order, top to bottom, with the first aligned to the bottom of $C_k$.
A scratch section can have one or more cells of any type: we extended cell creation to place a new cell in its parent container, be it a scratch section or the main notebook.

\mypara{Linear Execution}
\sys maintains Jupyter's cell-based interaction model, where cells are executed individually, but enforces a top-to-bottom execution order.
\newR{
Regardless of a cell $c$'s location or execution history, \sys ensures that executing $c$ yields a state consistent with the notebook's linear structure.    
To achieve this, when executing $c$, \sys clears the kernel's global namespace (via \scodewp{\%reset -f}) to remove effects from prior executions, computes $c$'s \emph{prefix} (defined below), and evaluates the code sequence \emph{prefix}+$c$ as the output for $c$.
}
\del{In other words}
\delR{When the user runs a new cell at the end of the notebook or a \spad section without skipping cells, termed \emph{linear}, the execution proceeds normally; when they rerun an existing cell $c$ (be it at the end or in the middle), termed \emph{nonlinear}, the execution is defined below.}

\delR{First, }\old{We compute the \emph{prefix} of $c$ }\delR{as follows}\newR{depending on its location}: 
\old{(1) if $c$ is in the notebook, \emph{prefix} is the code of all executed cells visually above $c$ concatenated in their visual ordering; (2) if $c$ is in the \spad, then $c$ must be in a scratch section $s$ attached to a notebook cell $c_{nb}$, and so \emph{prefix} is the concatenation of all executed notebook cells from the beginning of the notebook through $c_{nb}$, and all executed scratch cells in $s$ above $c$. 
Finally, \newR{with a kernel already reset, }we 
% \newR{reset the kernel namespace and }
run the code of $c$ prefixed by \emph{prefix}.}
\new
The above computation for \emph{prefix} \newR{along with kernel resetting} \new means that running a  cell in the main notebook \delR{, linearly or not,}\old will \emph{not} involve any code in the \spad, effectively sandboxing \spad execution from the notebook; similarly, running a cell in a given \spad section will \emph{not} involve any code from other \spad sections.
\old
This full-rerun approach is similar to the ``No Checkpoint'' approach in Multiverse~\cite{sato2024multiverse}. It involves no process forking and has reasonable performance (potentially faster than the serialization approach in Fork~It~\cite{weinmanForkItSupporting2021}; see Table 6 in Multiverse).

This rerun approach could cause code with side effects (\eg, printing) to repeat.
For example, if a cell $c$ contains a \scode{print} statement, and we run $c$ with the prefix \emph{prefix} that contains another \scode{print} statement, we get two print outputs on the cell instead of one; in regular Jupyter, we would have gotten only one output from $c$.
As a limited workaround, we modify \emph{prefix} as follows to avoid redundant output:
(1) we remove calls that display data to \scode{stdout}, including \scode{print} statements, plotting calls (\eg, \scode{plt.show()}), and dataframe summary calls (\eg, \scode{df.info()});
(2) we turn off \scode{matplotlib}'s interactive mode and enforce explicit calls of \scode{.show()} for plot display.

Note that our implementation of linear execution is less strict than the regular linear execution (in scripts and notebook systems like Polynote~\cite{Polynote}): we allow users to \emph{skip} cells, \newR{\ie, running a cell with one or more cells above it unrun.}\old

\mypara{\newR{Visualizing Stale Cells}}
\newR{Recall that the cell-based interaction model allows flexible execution orders, both (a) \emph{linearly}: running a new, previously-unrun cell at the end of the notebook or a \spad section with every cell above it already run in sequential order, and (b) \emph{nonlinearly}: everything else, including moving a cell from the \spad into the notebook.
In case of a nonlinear execution order upon cell $c$, all cells that come after (in the notebook or \spad) become \emph{stale}, and \sys informs the user as follows:}
\old
(1) clearing the execution results of $c$ (if any) and all cells occurring later in the notebook as well as their attached scratch sections, 
(2) regenerating results for $c$, and 
(3) graying out the later cells (in the notebook and \spad) to indicate their staleness.
\old

\mypara{Compatibility with Standard Jupyter}
\sys GUI information is written to the notebook metadata (JSON blob), including cell state (executed/folded/placement), code, \spad state (open/hidden), and scratch sections.
When a \sys notebook is opened in standard Jupyter, the main notebook cells come before the \spad cells,
and changes to a cell's content will not affect its placement (notebook/\spad) when re-opening the notebook in \sys.

\mypara{Generalizability and Limitations}
\sys implements always-clear notebook authoring on top of the classic Jupyter notebook (v6), but the implementation should extend to any computational notebook system that allows modifications to its GUI and the execution model. 
However, there are several limitations of our approach that we leave for future work.
First, our implementation of linear execution is incomplete as we only manually removed common calls that produce redundant output to \scode{stdout}.
Second, the implementation of linear execution
could be further improved with a more efficient, time traveling kernel~\cite{sato2024multiverse} or caching large computation to the disk~\cite{guo2011using}.

\section{Exploratory User Study}

We ran an IRB-approved exploratory user study to answer the following research questions: 

\aptLtoX{\begin{itemize}
    \item [\textbf{RQ1.}] {How do \sys features support maintaining clarity throughout notebook authoring?}
    \item [\textbf{RQ2.}] {How does \sys support realistic notebook tasks?}

    \item [\textbf{RQ3.}]{What strategies for notebook clarity does \sys enable?}
\end{itemize}}{\begin{itemize}
    \item [\RQ{1.}] {How do \sys features support maintaining clarity throughout notebook authoring?}
    \item [\RQ{2.}] {How does \sys support realistic notebook tasks?}

    \item [\RQ{3.}]{What strategies for notebook clarity does \sys enable?}
\end{itemize}}

\noindent
The full study materials are available in the supplements and at \url{https://tinyurl.com/tidynote-study-rep}.

\mypara{Participants}
We recruited 13 participants who were proficient with Jupyter notebooks via social media and emails.
\autoref{tab:participants} summarizes their demographics, frequency of Jupyter use, and attitude towards notebook clarity prior to \mbox{the study}.

\mypara{Tasks}
Since \sys aims to promote notebook clarity \emph{during} the development of notebooks, we choose to evaluate it with data analysis tasks, as they are often exploratory and prone to the accumulation of messes.
We provide a starter notebook for each participant.
The starter notebook imports one of the following datasets, both popular on Kaggle: ``Netflix Movies and TV Shows''\footnote{\url{https://www.kaggle.com/datasets/shivamb/netflix-shows/data}} or ``Most Streamed Spotify Songs 2023''\footnote{\url{https://www.kaggle.com/datasets/nelgiriyewithana/top-spotify-songs-2023/data}}.
The notebook continues with a tutorial of \sys in the context of data exploration.
Following the tutorial is a question about the given dataset to be answered with programming.
After the initial question, participants come up with their own question(s) about the dataset that they write code to answer; if they could not come up with any questions, we provide a list of questions for them to select.
As such, the tasks are open-ended: we do not restrict the answer to the initial question (\eg, participants could answer it with a plot, print statements, or something else, with or without library use) or the follow-up questions to resemble realistic data analysis work and to encourage in-depth usage of \sys.

\mypara{Procedure}
Each study took about 90 minutes on Zoom with informed consent.
Each participant was randomly assigned to 
one of the two starter notebooks (size of assignments balanced)
and accessed the notebook in the browser.
First, participants went through a \sys tutorial and practiced \sys features to familiarize with both the data and the environment.
Participants then spent up to 40 minutes (or until 20 minutes before the end of the study, whichever came first) writing code to answer the initial question provided and (if time permitted) their own questions about the dataset, with the ability to use Google search and AI assistants such as ChatGPT.
Participants were expected to keep the notebook as clear as possible, to the extent that they would be comfortable with continuing the analysis if leaving the notebook for weeks or sharing the work with collaborators.
For the remaining 20 minutes, participants \del{filled out a post-study survey and} \old went through a semi-structured interview to reflect on their experience.

\mypara{Data Collection}
From the pre-study screening survey, we collected participants' existing attitudes towards notebook clarity.
\del{From the post-study survey, we collected their ratings of \sys's usability with the Sytem Usability Scale (SUS), and their self-rated confidence in the clarity of the resulted notebooks as well as the effort in doing so via a five-point Likert scale.} \old
Throughout the study, we logged notebook actions (through \sys) and noted clarity-relevant behavior and quotes; the first author revisited notebook recordings to validate action logging.
Finally, the first author conducted open coding~\cite{clarke2017thematic} of the types of participants' own tasks from the final notebooks and field notes, as well as the post-study interviews (auto-transcribed by the conferencing software).

\newR{To measure how \sys helped maintain notebook clarity during realistic tasks, we collected the final notebooks created by the participants and evaluated the importance of each Python code statement (\eg, assignment, function definition, import) to their analysis tasks. 
Though there is no standard rubric for such evaluation, following practitioners' definitions for notebook cleaning of ``keeping a desired subset of results while discarding
the rest''~\cite{Head2019:Managing}, the first two authors created three categories for the code statements in the final notebooks: (1) \emph{Relevant} if directly generating results for the analysis tasks; (2) \emph{Necessary} if not relevant but syntactically necessary for the code to run (\eg, function definitions and library imports); (3) \emph{Transient} if neither relevant nor necessary, such as one-off exploration.
The first author reviewed all final notebooks to categorize all statements after the tutorial portion.
\autoref{tab:tidiness-eval} reports the results of this analysis, which we discuss in~\autoref{subsec:realistic-tasks}.
}

\mypara{Study Limitations}
First, although we recruited participants from various backgrounds, there were only 13 participants, most of whom worked in computing-relevant fields with self-reported Jupyter experience.
Second, although participants worked on their own analyses of the given data that could improve external validity, our findings were limited to data-driven notebook work and the two datasets involved.
Third, there was no explicit comparison between \sys and a baseline system\newR{, as we decided from the comparative analysis (\autoref{subsec:comparative-analysis}) that no prior system was intended for maintaining notebook clarity throughout the authoring process, \ie, the goal of \sys}.
\newR{To mitigate this, because participants were familiar with the regular Jupyter, we asked them to retroactively compare their \sys experience to prior Jupyter experience in the post-study interview.}
\new
Finally, our study required participants to keep their notebooks as clear as possible to stress-test \sys features as opposed to assessing its most natural use.

As such, our study findings should be viewed as an early step towards understanding the effects of always-clear \del{computational notebooks} \new{notebook authoring} and generalizations to notebook systems beyond Jupyter.

\aptLtoX{\begin{table*}[t]
\centering
\caption[]{Left: Participant demographics, occupation, Jupyter use, and attitude towards notebook clarity prior to the study (``Reluctant''=would reluctantly clean a notebook, ``Glad''=would gladly clean a notebook, ``Always''=would always keep a notebook clear. Right: Number of tasks beyond the provided task performed during the study, and task types.}
\label{tab:participants}
\begin{tabular}{llll|p{6.5cm}}
\toprule
 & \textbf{Occupation} & \textbf{\makecell[l]{Jupyter\\Use}} & \textbf{\makecell[l]{Clarity\\Preference}} & \textbf{Num ``Other'' Tasks and Task Types}   \\
\midrule
P1 (M) & PhD student, systems  & Daily & Reluctant & 5: wrangling (2); summarizing (2); plotting (1)  \\
P2 (M) & PhD student, AI/ML & Daily & Reluctant & 1: modeling (1)  \\
P3 (M) & PhD student, security & Daily & Always & 2: plotting (2)  \\
P4 (M) & Software developer & Daily & Always & 0: N/A  \\
P5 (M) & Professor, Chemistry & Daily & Reluctant & 1: wrangling (1) \\
P6 (M) & ML engineer, finance & Daily & Glad & 4: summarizing (4) \\
P7 (M) & ML engineer & Daily & Always & 7: summarizing (6); filtering (1)  \\
P8 (M) & PhD student, IoT & Daily & Reluctant & 1: plotting (1)  \\
P9 (F) & Student, data science & Regularly & Always & 1: wrangling (1)  \\
P10 (F) & Student, AI/ML  & Regularly & Always & 1: plotting (1)  \\
P11 (F) & Researcher, CS education & Occasionally & Reluctant & 2: language processing (1); filtering (1) \\
P12 (F) & Student, computer science & Occasionally & Always & 3: summarizing (1); plotting (2) \\
P13 (F) & Hydrologist, geoscience & Regularly & Glad & 2: summarizing (1); wrangling and plotting (1) \\
\bottomrule
\end{tabular}
\Description{Table listing all participants, their genders, occupations, Jupyter use frequency, preference for clarity prior to the study, and the number of their own tasks and task types during the study.
}
\end{table*}}{\begin{table*}[t]
\centering
\caption[]{\textbf{Left}: Participant demographics, occupation, Jupyter use, and attitude towards notebook clarity prior to the study (``Reluctant''=would reluctantly clean a notebook, ``Glad''=would gladly clean a notebook, ``Always''=would always keep a notebook clear. \textbf{Right}: Number of tasks beyond the provided task performed during the study, and task types.}
\label{tab:participants}
\begin{tabular}{llll|p{6.5cm}}
\toprule
 & \textbf{Occupation} & \textbf{\makecell[l]{Jupyter\\Use}} & \textbf{\makecell[l]{Clarity\\Preference}} & \textbf{Num ``Other'' Tasks and Task Types}   \\
\midrule
P1 (M) & PhD student, systems  & Daily & Reluctant & 5: wrangling (2); summarizing (2); plotting (1)  \\
P2 (M) & PhD student, AI/ML & Daily & Reluctant & 1: modeling (1)  \\
P3 (M) & PhD student, security & Daily & Always & 2: plotting (2)  \\
P4 (M) & Software developer & Daily & Always & 0: N/A  \\
P5 (M) & Professor, Chemistry & Daily & Reluctant & 1: wrangling (1) \\
P6 (M) & ML engineer, finance & Daily & Glad & 4: summarizing (4) \\
P7 (M) & ML engineer & Daily & Always & 7: summarizing (6); filtering (1)  \\
P8 (M) & PhD student, IoT & Daily & Reluctant & 1: plotting (1)  \\
P9 (F) & Student, data science & Regularly & Always & 1: wrangling (1)  \\
P10 (F) & Student, AI/ML  & Regularly & Always & 1: plotting (1)  \\
P11 (F) & Researcher, CS education & Occasionally & Reluctant & 2: language processing (1); filtering (1) \\
P12 (F) & Student, computer science & Occasionally & Always & 3: summarizing (1); plotting (2) \\
P13 (F) & Hydrologist, geoscience & Regularly & Glad & 2: summarizing (1); wrangling and plotting (1) \\
\bottomrule
\end{tabular}
\Description{Table listing all participants, their genders, occupations, Jupyter use frequency, preference for clarity prior to the study, and the number of their own tasks and task types during the study.}
\end{table*}}

\section{Results}\label{sec:results}

Below we report the user study results according to our three RQs in terms of behavioral data, quotes, and comparison to regular Jupyter.

\begin{figure*}[htbp]
  \centering
  \includegraphics[width=\textwidth]{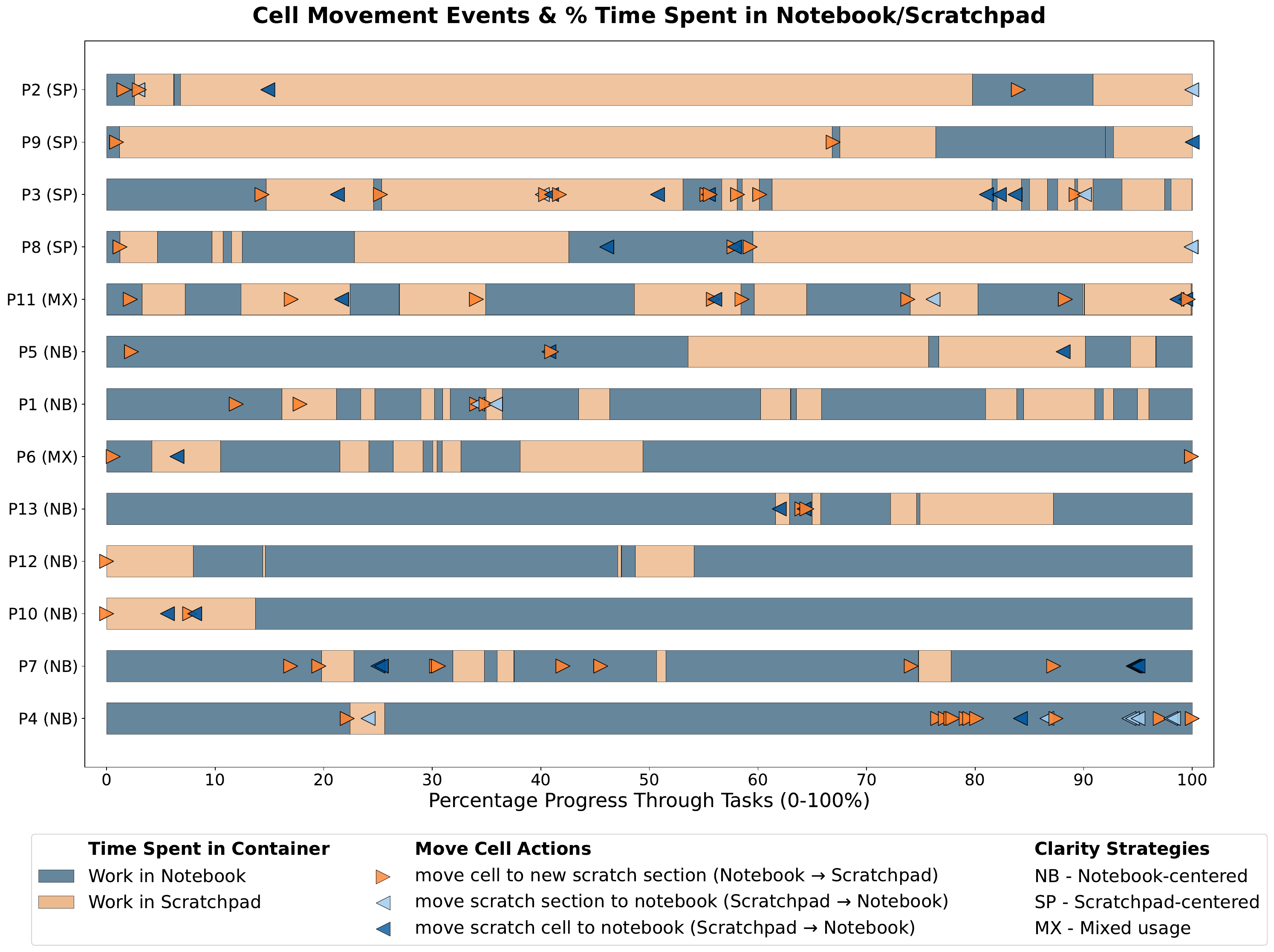}
  \caption{Cell movement events \& percentage time spent in notebook/scratchpad, ordered by \% time spent in \spad from top to bottom. ``Time through tasks'' is the interval between the onset of the first notebook action after the tutorial, and the ending of the last action before the end of the tasks. Time progress is normalized across participants, labeled by their strategies for clarity (\autoref{subsec:results-strategies}).}
  \label{fig:merged-temporal-analysis}
  \Description{A horizontal stacked bar chart titled "Cell Movement Events & \% Time Spent in Notebook/Scratchpad" displaying the work of thirteen participants over the course of a task. The x-axis represents the "Percentage Progress Through Tasks" from 0 to 100. Each participant's work is shown as a horizontal bar, color-coded to indicate the time spent in container: dark blue for "Work in Notebook" and light orange for "Work in Scratchpad". The y-axis represents the participants.
  Participants are ordered by their \% spent in the scratchpad, from the most to the least, and from top to bottom: P2, P9, P3, P8, P11, P5, P1, P6, P13, P12, P10, P7, P4.  
  Small triangles on the bars denote "Move Cell Actions" , such as moving a cell from the notebook to the scratchpad (orange, right-pointing triangle) or from the scratchpad to the notebook (left-pointing triangles, dark blue for moving a cell, light blue for moving a scratch section).
  Participant labels on the y-axis are annotated by their "Clarity Strategies": "SP - Scratchpad-centered" (P2, P9, P3, P8), "MX - Mixed usage" (P11, P6), and "NB - Notebook-centered" (P5, P1, P13, P12, P10, P7, P4). Definitions of these strategies in Section 7.2.}
\end{figure*}

\subsection{RQ1: \new{How do \sys features support maintaining clarity throughout notebook authoring?} \old
}\label{subsec:system-usage}

We describe how participants in our study used various features of \sys that achieve our three Design Goals (DGs)
\new{to maintain notebook clarity. }\old
First, we report the use of features that implement \textit{\DGOne}.

\begin{figure*}[t]
  \centering
  \includegraphics[width=.95\textwidth]{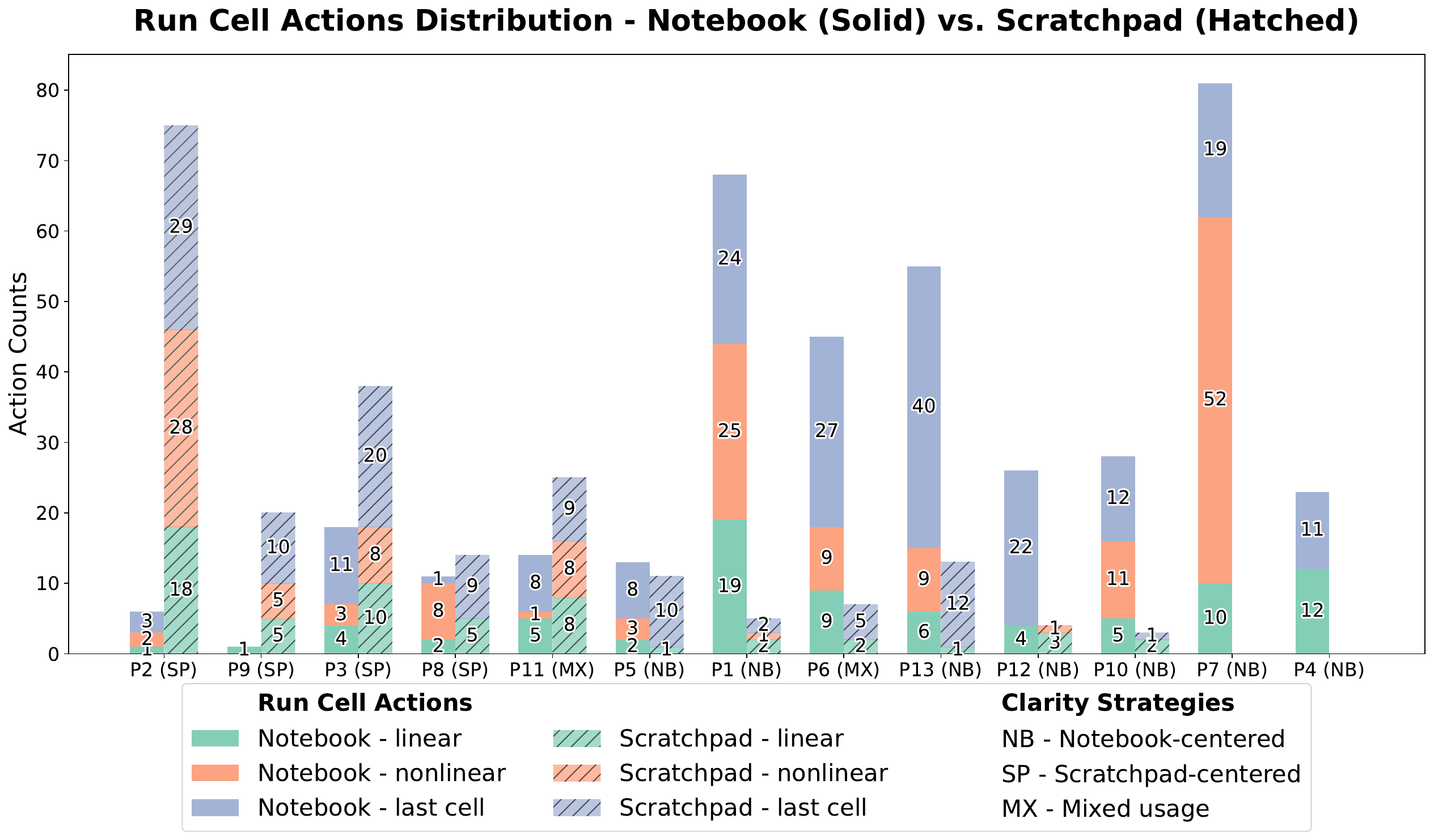}
  \caption{Distribution of run cell actions by container. ``linear'' = running a new\newR{, previously unrun cell at the end of the notebook or a \spad section with all cells above it already run in sequence, }\old ``last cell'' = rerunning the (previously-run) cell at the end of the notebook or a \spad section (cells above it already run in sequence), ``nonlinear'' = everything else. Participants are labeled with their strategies for clarity (\protect\autoref{subsec:results-strategies}) and ordered the same as \protect\autoref{fig:merged-temporal-analysis}.}
  \label{fig:merged-actions-distribution}
  \Description{A stacked bar chart titled "Run Cell Actions Distribution - Notebook (Solid) vs. Scratchpad (Hatched)" The chart displays the number of cell executions for thirteen participants, who are listed along the x-axis and labeled by their "Clarity Strategy" - NB: Notebook-centered, SP: Scratchpad-centered, MX: Mixed usage, as defined in Section 7.2.
  Participants are ordered by their \% spent in the scratchpad, from the most to the least, and from left to right: P2 (SP), P9 (SP), P3 (SP), P8 (SP), P11 (MX), P5 (NB), P1 (NB), P6 (MX), P13 (NB), P12 (NB), P10 (NB), P7 (NB), P4 (NB).
  The y-axis shows the "Action Counts".
  Each participant is represented by a stacked bar, where each segment's color and pattern indicate the type of action.
  Solid colors represent actions in the Notebook, while hatched patterns signify actions in the Scratchpad.
  Green denotes "linear" execution, orange denotes "nonlinear" execution, and blue denotes executing the "last cell".
  Each colored segment is labeled with the specific count for that action type, given the order of Notebook - linear, Notebook - nonlinear, Notebook - last cell, Scratchpad - linear, Scratchpad - nonlinear, Scratchpad - last cell:
  P2 - 1,2,3,18,28,29;
  P9 - 1,0,0,5,5,10;
  P3 - 4,3,11,10,8,20;
  P8 - 2,8,1,5,0,9;
  P11 - 5,1,8,8,8,9;
  P5 - 2,3,8,1,0,10;
  P1 - 19,25,24,2,1,2;
  P6 - 9,9,27,2,0,5;
  P13 - 6,9,40,1,0,12;
  P12 - 4,0,22,3,1,0;
  P10 - 5,11,12,2,0,1;
  P7 - 10,52,19,0,0,0;
  P4 - 12,0,11,0,0,0.
  The chart illustrates that participants with a Scratchpad-centered (SP) strategy, like P2, tend to have more cell execution actions in the scratchpad (hatched segments), while Notebook-centered (NB) participants, like P7, have more actions in the notebook (solid segments).}
\end{figure*}

\mypara{[DG1] All participants alternated notebook and \spad work\new{ to structure exploration}}
In \sys, one could \emph{actively work} in the notebook or the \spad by creating, running, or (vertically) moving cells. 
We identified \emph{container switch} events---when the locations of two consecutive actions differ---and computed the proportions of time each participant actively worked in the notebook/\spad.
The colored blocks in \autoref{fig:merged-temporal-analysis} show the data, participants are ordered from top to bottom by the percentage of active work in the \spad.
The figure shows that all participants actively worked in both containers, and that they all alternated their active work between the containers throughout the tasks.
In addition to alternating between containers, four participants (P1, P3, P7, P11) further alternated between at least two scratch sections in the \spad, using the additional sections for reference (P1, P7), one-off explorations (P3, P11), reminders for future tasks (P3), and archival of outdated or irrelevant code (P7).

\mypara{[DG1] Most participants pinned at least one cell to minimize scrolling during exploration}
11 out of 13 participants (all but P1, P2) used cell pinning to keep inspection results or reference material visible as they sifted through long notebooks. 
During the post-study interview, P6 further showed a personal notebook to suggest pinning non-code cells like markdown checklists, something the study did not demonstrate but \sys already supports.
Two participants (P1, P2) did not pin any cell, either feeling the lack of need with all the work within the viewport (P2) or having forgotten about it (P1).
P1 recognized this as a \emph{``user error''}, but suggested \emph{``automatically pinning [in] the \spad''} as he had to scroll repeatedly through the \spad during the study.

\smallskip\noindent
Below is the use of features relevant to \textit{\DGTwo}.

\mypara{[DG2] All participants moved cells between the notebook and \spad throughout the study\new{ as exploration progressed into clarity}}
\autoref{fig:merged-temporal-analysis} also visualizes the cell movement events as triangles in their temporal order, showing that all participants moved cells at least once throughout their tasks.
The \spad is designed to be used in conjunction with the notebook: one can create new cells in existing scratch sections, but can only create a new scratch section by moving a notebook cell over.
Even with this requirement, all participants except P12 moved cells in both directions.
Most participants except for P8 found moving cells intuitive.
P8 initially struggled with restoring the ordering of \spad cells in the notebook, but later realized that moving an entire \spad section was meant for moving a group of cells.
Indeed, six participants (P1, P2, P3, P4, P8, P13) moved entire \spad sections to the notebook when they wanted to include entire sections of scratch work in the main notebook narrative.

\mypara{[DG2] Some participants toggled the \spad to focus on the notebook or confirm notebook clarity}
Eight participants (P1, P2, P3, P7, P8, P10, P11, P13) toggled the \spad at least once during the study.
Most of them hid the \spad to temporarily focus on the notebook (P1, P2, P3, P7, P10, P11, P13), and three participants (P8, P10, P11) explicitly closed the \spad at the end of all the tasks to read through the notebook and ensure it was clear and presentable.
Among the participants who did not toggle the \spad, they either found it unnecessary to hide it for purposes other than presentation (P5, P6, P9) or they misinterpreted the button as deleting the \spad (P4, P12); P12 \emph{``would definitely use that if [she] knew that next time.''}

\smallskip\noindent
Finally, we report on the use of linear execution for \textit{\DGThree}.

\mypara{[DG3] All participants ran cells out of order without experiencing confusion with state}
\sys enforces linear execution, a big deviation from the regular Jupyter semantics that allow execution in any order.
We analyzed  how linearity affected (a) participants' execution behavior and (b) state clarity.
For (a), we recorded three types of cell execution: (1) ``linear''---running a new\newR{, previously unrun cell at the end of the notebook or a \spad section with all cells above it already run in sequence }\old (2) ``last cell''---rerunning the cell at the end of the notebook or a \spad section (cells above it already run in sequence) (3) ``nonlinear''---every run that does not fall under the previous categories.
\autoref{fig:merged-actions-distribution} visualizes the distribution of these actions, showing that out-of-order execution (``nonlinear'' and ``last cell'') remained prevalent regardless of where the execution occurred (notebook or \spad).
For (b), we measured how participants used Jupyter's ``restart and run all'' feature, a common technique to restore state clarity by restarting the kernel and running all cells~\cite{huang2025howscientists}.
We found that \emph{no participant} ever had to use this feature during the study to resolve state confusion; only P2 did so intentionally to confirm his notebook was reproducible at the end of the tasks.
In addition, although we did not ask explicitly, no participant mentioned confusion about state, and no participant had to manually trace/debug the state.
Finally, we reran copies of the 13 notebooks used in the study as a sanity check, finding that all of them reproduced the same results as seen in the study.

\mypara{Comparison to Regular Jupyter}
Overall, participants recognized that \sys features work together to improve the experience of flexible exploration and maintaining clear notebooks throughout the notebook authoring workflow. 
As P7 noted, \emph{``I don't think any of these features are unnecessary---it just depends on which use case you're working on.''}
Participants commented on the following improvements compared to regular Jupyter:
\begin{itemize}

    \item \textbf{\Spad and movable cells encourage exploration without compromising notebook clarity}. The \spad enables temporary archival of prior exploration or messy code that could be useful in the future (P1, P3, P4, P7, P8), which if created in the notebook can be easily moved out (or back) to keep the notebook clear. 
    In addition, the \spad separates exploration and clutter (P2, P3, P5, P9, P10, P11, P12, P13), supports comparative data exploration (P2, P3, P9, P12, P13), and eases hypothesis testing (P5, P11) such as validating AI-generated results before incorporating them (P11). 
    All of the above are not available in regular Jupyter.

    \item \textbf{Cell pinning eases navigation during exploration}.
    P10, who pinned the most cells among all participants (five), explicitly called out its benefit for avoiding scrolling, while she \emph{``would have had more redundant cells to avoid the scrolling''} in regular Jupyter.
    Scrolling is indeed the activity with the most time spent in regular Jupyter~\cite{huang2025howscientists}, and as P3 put it, \emph{``I've died so many times scrolling around my notebook and trying to find [something]. [Pinning] is perfect.''}
    Six participants (P3, P6, P7, P8, P10, P12)  highlighted how pinning reduced the need to repeat inspection code like \scode{df.head()} at multiple points.

    \item \textbf{Linear execution ensures clarity of program flow} by maintaining a clear and consistent program state, which was especially helpful for workflows like machine learning. 
    As P6 demonstrated in his personal Jupyter notebook at several places, forgetting to re-run key steps caused errors. 
    In addition, 10 participants (P1, P3, P4, P5, P6, P7, P10, P11, P12, P13) thought that linearity helped avoid confusion and encourage regular testing, and consequently led to less error-prone notebooks.

    \item \textbf{\sys can enhance teaching with notebooks}, an opportunity recognized by two participants with relevant teaching experience (P5, P11).
    They envisioned live coding~\cite{selvaraj2021live} or peer instruction~\cite{crouch2001peer} scenarios, where both the instructor and the students could experiment in the notebook
    while keeping it \emph{``nicely structured''} (P5).
    P11 liked how \sys supports one-off explorations without compromising notebook clarity: she taught in scenarios where students sometimes did not have Internet access and could not explore relevant concepts on their own, so they relied on her to demonstrate the ``what-if'' scenarios in the notebook used for lecture.

\end{itemize}

Still, some participants felt \sys features did not change or even worsened certain scenarios that did not require clarity, such as using a notebook entirely for scratch work (P8, P9) and small tasks that might eventually be scripted (P9).
For these contexts, the enforced linearity limits the ability to explore freely (P8, P9) and compromises performance when repeatedly importing large datasets (P8), and the \spad could have been just a sandbox independent from the notebook as opposed to having sections attached to different places (P9).

\subsection{\new{RQ2: How does \sys support realistic notebook tasks?}}\label{subsec:realistic-tasks}

\begin{table}[t]
\small
\centering
\caption[]{\tidinessEvalCaption}
\setlength{\tabcolsep}{1em} 
\begin{tabular}{c|ccc|ccc}
\toprule
& \multicolumn{3}{c|}{\thead{\customnormal \textbf{Notebook}}} & \multicolumn{3}{c}{\thead{\customnormal \textbf{Scratchpad}}} \\
\thead{\customheader \textbf{ID}} & \thead{\customheader \textbf{R}} & \thead{\customheader \textbf{N}} & \thead{\customheader \textbf{T}} & \thead{\customheader \textbf{R}} & \thead{\customheader \textbf{N}} & \thead{\customheader \textbf{T}} \\
\midrule
P1 & 10 & 10 & 2 & 0 & 0 & 1 \\
P2 & 25 & 44 & 0 & 0 & 0 & 0 \\
P3 & 12 & 14 & 1 & 0 & 0 & 0 \\
P4 & 6 & 8 & 0 & 0 & 0 & 3 \\
P5 & 2 & 2 & 0 & 0 & 0 & 1 \\
P6 & 6 & 2 & 1 & 0 & 0 & 2 \\
P7 & 9 & 4 & 2 & 0 & 0 & 0 \\
P8 & 6 & 9 & 1 & 0 & 0 & 0 \\
P9 & 2 & 5 & 0 & 0 & 0 & 1 \\
P10 & 9 & 14 & 0 & 0 & 0 & 0 \\
P11 & 6 & 10 & 1 & 0 & 0 & 10 \\
P12 & 4 & 3 & 0 & 0 & 0 & 0 \\
P13 & 15 & 11 & 3 & 0 & 0 & 5 \\
\midrule
\textit{Min.} & \textit{2} & \textit{2} & \textit{0} & \textit{0} & \textit{0} & \textit{0} \\
\textit{Mean} & \textit{8.6} & \textit{10.5} & \textit{0.8} & \textit{0.0} & \textit{0.0} & \textit{1.8} \\
\textit{Max.} & \textit{25} & \textit{44} & \textit{3} & \textit{0} & \textit{0} & \textit{10} \\
\bottomrule
\end{tabular}
\Description{\tidinessEvalDescription}
\label{tab:tidiness-eval}
\end{table}

\new{\autoref{tab:participants} (column ``Num Other Tasks and Task Types'') shows that, during the study, \sys supported participants in six types of data work: wrangling (manipulating data in place), summarizing (using library functions to summarize data without manipulation), plotting (using libraries to plot data), modeling (creating models to predict unseen data based on the given data), filtering (obtaining a subset of data based on conditions), and language processing (analyzing patterns in natural language data).
All participants felt that their tasks during the study resembled their usual notebook tasks.
In addition to the tasks performed during the study, several participants imagined using \sys to teach intro programming (P5, P11) and document longitudinal findings (P3, P8, P13).}

\newR{Our analysis of the code statements in the participants' final notebooks (\autoref{tab:tidiness-eval}) further suggests that \sys helped participants maintain clear notebooks.
On average, the notebook portion had 8.6 \emph{Relevant} statements (assignments, function definitions, etc.) and 10.5 \emph{Necessary} statements, compared to only 0.8 \emph{Transient} statements.
In contrast, the \spad portion had zero \emph{Relevant} and \emph{Necessary} statements, and 1.8 \emph{Transient} statements, across all participants.
This indicates that the \emph{Transient} code that did not get deleted along the way mostly ended up in the \spad, suggesting that the resulting notebook portion could be considered more narrative focused and thus more ``tidy,'' while the \spad contained extraneous computation.
Moreover, six participants had no code left in the \spad in their final notebooks, either deleting them all (P2, P3, P8, P12) or moving them back into the notebook (P7, P10) before the end of their tasks. 
The notebook and \spad separation may have assisted this final cleaning:
had those cells been in the notebook, these participants would have had a harder time deciding which cells were deletable or movable.
}

\new{%
\mypara{Comparison to Regular Jupyter}
In addition to benefiting from the \sys features, most participants appreciated the ability to reuse familiar Jupyter workflows in \sys, which made it learnable and adaptable to regular notebook tasks.
Specifically, while participants felt \emph{what} they did in the study resembled their usual work, when asked to compare \emph{how} they did the tasks during the study with their usual Jupyter workflow,
12 participants (all except P9) felt that their workflows with \sys were similar to their usual workflows.
Indeed, in addition to the \sys features aimed for always-clear authoring, participants had access to all standard Jupyter features in \sys.
P4 particularly called out that \sys's tight integration into Jupyter eased adjusting to the new interaction paradigm because most prior Jupyter usage patterns remained available.}\old

\subsection{\new{RQ3: What strategies for notebook clarity does \sys enable?}}\label{subsec:results-strategies}
To understand how participants used \sys to maintain clarity throughout notebook authoring, we revisited their notebook activities from the video recordings and instrumentation data.
The analysis showed two main strategies: (1) a \emph{notebook-centered strategy}, if the user keeps their work within the main notebook in \sys and wraps up a task by deleting the messy cells or moving them to the \spad; and (2) a \emph{\spad-centered strategy}, if the user keeps their work within the \spad and only moves it back into the main notebook when wrapping up.

 \mypara{Notebook-centered}
Seven participants (P1, P4, P5, P7, P10, P12, P13) adopted the notebook-centered strategy to keep their notebook clear.
Instrumentation data shows that all these participants spent a greater portion of their time in the notebook than in the \spad, creating and executing more cells in the notebook, as is illustrated in both \autoref{fig:merged-temporal-analysis} (more blocks in blue) and \autoref{fig:merged-actions-distribution} (higher solid bars).
Notably, P4 and P7 never ran cells in the \spad (no corresponding hatched bars in \autoref{fig:merged-actions-distribution}), only moving cells back and forth as needed (\autoref{fig:merged-temporal-analysis}).

 \mypara{\Spad-centered}
Four participants (P2, P3, P8, P9) used the \spad-centered strategy, spending more time creating and running cells in the \spad (top four rows with more orange blocks in \autoref{fig:merged-temporal-analysis}, and higher hatched bars in \autoref{fig:merged-actions-distribution}).
Two participants (P2, P9) \emph{cherry-picked} their work from the \spad: P2 manually copied and pasted code line by line into the main notebook; P9 cherry-picked similarly but in a new cell within a scratch section before eventually moving the cell back into the main notebook (few blue triangles for P2 and P9 in \autoref{fig:merged-temporal-analysis}, indicating fewer cell movements).
P3 did not cherry-pick by copy-paste but instead directly moved individual cells and scratch sections back into the notebook (more blue triangles than P2, P9, and P8 in \autoref{fig:merged-temporal-analysis}).
P8 adopted the P2's cherry-picking strategy in the initial task and P3's strategy of direct cell movement in his own task.

 \mypara{Mixed usage}
Two participants (P6, P11) used a mix of both strategies.
P6 started his work with the \spad-centered strategy for the initial task but eventually leaned toward the notebook-centered way (more orange blocks in the beginning than later in \autoref{fig:merged-temporal-analysis}); as he worked on four additional tasks (\autoref{tab:participants}), he ran more cells in the main notebook (higher solid bar in \autoref{fig:merged-actions-distribution}).
P11, in contrast, started her work in a notebook-centered way but gradually moved data exploration and hypothesis testing into the \spad  before moving the concluding results back into the notebook, and she indeed ran cells more in the \spad (higher hatched bar in \autoref{fig:merged-actions-distribution}).

\mypara{Comparison to Regular Jupyter}
In the interview, we asked all participants how \sys affected their attitudes towards notebook clarity.
With the above \sys-inspired strategies for notebook clarity, \emph{all} participants who were previously unmotivated about clarity (\autoref{tab:participants}: P1, P2, P5, P6, P8, P11, P13) felt that clarity could be easily achieved in \sys within their preferred strategy.
Because \sys \emph{``doesn't really introduce that much overhead [of cleaning] in [one's] workflow''} (P1), P2's workflow was completely changed: \emph{``Right now, I would clean my notebook. Before, I would not clean my notebook.''}
Before with regular Jupyter, participants delayed cleanup until necessary (P6) with manual post-hoc cleaning (P6, P10) or help from AI assistants (P1, P2), or simply ignored notebook clarity (P1, P2, P11).
Meanwhile, those already committed to notebook clarity (P3, P4, P7, P9, P10, P12) maintained the attitude after using \sys,
although P9 would rather not use \sys for clarity out of her distaste for enforced linearity.
Finally, there are cleaning workflows that participants would reuse from prior Jupyter experience, such as adding manual documentation and markdown headings (P3, P13).

%%%%%
\section{Discussion and Future Work}

Based on our study results, below we discuss how \sys meets its design intention (\autoref{subsec:discussion-goal}), implications for clarity workflows in regular Jupyter (\autoref{subsec:discussion-generalizability}), design opportunities for future notebook clarity support (\autoref{subsec:discussion-clarity-def}-\autoref{subsec:discussion-clarity-side-effects}), \new{and design implications for future information systems (\autoref{subsec:discussion-sharing}).}\old

\subsection{Always-Clear Authoring: Does \sys support it?}\label{subsec:discussion-goal}
Our main goal is to support always-clear notebook authoring, by meeting our three Design Goals to better manage exploration (DG1), enable rapid iteration between exploratory and non-exploratory activities (DG2), and promote clarity in program state (DG3).
In \sys, we proposed the \spad and cell pinning to support DG1, cells movable between the notebook and \spad to support DG2, and linear execution with state branching in the \spad to support DG3.
Our study showed that participants used all these features interchangeably \new{to maintain clarity throughout a notebook's lifecycle }\old(\autoref{subsec:system-usage})\new{ for realistic notebook tasks (\autoref{subsec:realistic-tasks})}\old, \new{and }\old developed strategies that suited their workflows for maintaining clarity (\autoref{subsec:results-strategies}).
% , and considered \sys to be functionally usable (\autoref{subsec:results-usability}).
These results highlighted that \sys is a promising step towards supporting always-clear authoring. 

There are two possible explanations for the generally positive findings.
First, we grounded the design in empirical findings about notebook use and a thorough comparative analysis of existing tools.
Second, instead of building a brand new system, we implemented an augmentation of Jupyter largely compatible with existing features.
Although a further discussion on interface learnability is beyond the scope of this work, prior work assessing the learnability of program synthesizers~\cite{jayagopalExploringLearnabilityProgram2022} highlighted the importance of supporting pre-existing experience.
As a Jupyter extension, \sys allows participants to leverage pre-existing knowledge about Jupyter notebooks (\autoref{subsec:realistic-tasks}). And \sys's linear execution, though different from Jupyter, is close to the regular linear execution in scripting, where code is executed top to bottom.

\subsection{Generalizability of Findings}\label{subsec:discussion-generalizability}
To stress-test \sys features for maintaining notebook clarity, our study required participants to keep their notebooks as clear as possible.
This may differ from notebook authoring in the real world where clarity may not be enforced.
Our results suggested that, once leaving the \sys bubble, participants would likely resume their prior clarity habits: those reluctant about maintaining clarity (with regular Jupyter) only felt more motivated about clarity during the study because \sys eased achieving the goal (\autoref{subsec:results-strategies}).

\begin{table*}[t]
  \small
%   \setstretch{1.25}
  \centering
  \caption[]{User Tactics for Notebook Clarity Supported by \sys}
  \label{tab:tactics}
  \begin{tabular}{c|c|c}
    \hline
    \textbf{\ \ Clarity Requirement\ \ } & \textbf{Manual Tactics in Jupyter}~\protect\cite{huang2025howscientists} & \textbf{\ \ \sys Support\ \ } \\
    \hline
    \multirow{4}{*}{Content} & Removing/commenting out redundant code and cells & \multirow{3}{*}{\Spad} \\
    \cline{2-2}
    & Using empty cells to separate groups of cells & \\
    \cline{2-2}
    & \ Debugging in a fresh notebook to keep the original intact\  & \\
    \cline{2-3}
    & Merging cells relevant to one task & Scratch sections \\
    \hline
    State & Using abstractions (\eg, function definitions) & Linear execution \\
    \hline
  \end{tabular}
  \Description{Table of 5 rows and 3 columns enumerating Clarity Requirement, Manual Tactics in Jupyter as reported in prior work (citation 15), and corresponding Tidynote support. 
  Rows 1-4 has a few cells spanning over multiple rows: Rows 1-4 all have clarity requirement of Content, with four tactics reported in each row; Rows 1-3 have Tidynote support of Scratchpad, while Row 4 has support of scratch sections.}
\end{table*}

Still, the study results and \sys remain relevant to practical notebook authoring.
A meaningful portion of notebook users genuinely value clarity but have to rely on manual tactics to meet their goals in regular Jupyter. Six out of 13 participants in our study always kept clear notebooks in their ordinary work (\autoref{tab:participants}). Moreover, an earlier study found that 17 out of 20 participants adopted tactics to clean their notebooks, and some of these participants cleaned continuously~\cite{huang2025howscientists}.
But, cleaning is only indirectly supported in ordinary Jupyter, so many of the user tactics have more of the character of ``workarounds'' rather than ``using Jupyter features as intended''.
\autoref{tab:tactics} shows several of these previously reported manual tactics for clarity~\cite{huang2025howscientists}, as well the corresponding features in \sys that streamline or subsume the tactic.
For example, during the study P11 leveraged \spad's support for the debugging (third tactic in \autoref{tab:tactics}), which inspired part of our walkthrough example in \autoref{sec:walkthrough}.
The tactic ``merging cells relevant to one task'' (fourth tactic in \autoref{tab:tactics}) comes at the cost of losing intermediate cell outputs~\cite{huang2025howscientists}, which \spad sections avoid by allowing relevant cells to be in a section.
Abstracting code into a function to avoid polluting the global state (last tactic in \autoref{tab:tactics}), in addition to requiring a tedious refactor, also sacrifices the ability to display intermediate outputs within the function for exploration and debugging; in comparison, \sys's linear execution keeps the global state cleaner, potentially delaying the need to abstract code into a function.

\subsection{\del{Clarity: A Standard or Spectrum?}\new{A Lack of Standard for Clarity}}\label{subsec:discussion-clarity-def}
Our study also revealed an interesting contrast between some participants' own feature preferences and the resultant clarity of their notebooks.
% A notable example is the difference between P2 and P9.
A notable example is the difference between P2 and P9: P9 disliked linearity but favored the \spad, while P2 was the opposite.
% In terms of preference, P9 disliked linearity but favored the \spad, while P2 was the opposite.
% \todo{Why does it matter that they could have done things differently for themselves? Is the point simply that they did work differently from each other and yet still rated their notebooks clear? And how does that motivate "a bigger issue with notebook clarity"?}
% Following their preferences, P9 could have run cells only linearly to avoid rerunning stale cells caused by enforced linearity, and P2 could have avoided using the \spad entirely.
% On the contrary, P9 still ran more cells nonlinearly, and P2 still spent most of his time working in the \spad.
Despite the contrasting feature preferences and resultant differences in \sys usage, in the post-study interview, they both felt highly confident with their notebook clarity\newR{, and their notebooks (the notebook portion) contained entirely \emph{Relevant} and \emph{Necessary} code, according to our evaluation (\autoref{tab:tidiness-eval})}\old.
These results suggest that the interpretation of ``clarity'' could be subjective: \eg, clarity in state was important to P2 but not P9.

% These results provided additional evidence for how \sys features worked collectively to meet its design goals
% These results revealed a bigger issue with notebook clarity: 
Indeed, there is no standard of what makes a notebook ``clear''.
% \newR{, and even our \emph{tidiness} evaluation was an approximation based on prior work~\cite{Head2019:Managing} and our own understanding}\old. 
Notebooks lack something similar to software development guidelines despite empirical results on how notebook users maintain clarity~\cite{weinmanForkItSupporting2021, huang2025howscientists}.
% We attribute the observations to the lack of a standard for notebook clarity similar to software development guidelines despite empirical results on how notebook users maintain clarity~\cite{weinmanForkItSupporting2021, huang2025howscientists}.
We observed varying standards for clarity among participants: 
% As P8 commented, given his typically unstructured note-taking and exploration habits in notebooks, he lacked a reference for a ``clear'' notebook.
P4 thought that code readability was the top criterion for notebook clarity;
P8 had a vague concept for clarity as he typically did not keep a ``clear'' notebook;
others (\eg, P3 and P13) valued the ability to maintain a cohesive narrative more.
\sys does not directly account for the varying standards for clarity, although it still caters to different needs by providing multiple features aimed for different aspects of clarity (\eg, content vs. state), as illustrated by the example of P2 and P9 above.
Given the variety of views and strategies for notebook clarity~
(\autoref{subsec:results-strategies}), we believe that clarity should be judged relative to the user's task and preference, and future always-clear notebook authoring systems should provide support for multiple aspects of clarity, adequate for the user's desired level and preference.
% and kind of clarity.
% \todo{Does \sys somehow not? Might be able to say it kinda does support the different needs...like you said above: "Despite the contrast...they gave the highest rating". If we are going to suggest something about future systems, ideally we'd suggest a concrete idea too}

% and upon the their own discretion.
% \emph{\sys supports users to meet their own standard for clarity in their preferred way, and future always-clear notebooks should maintain such flexible support.}

%\lisa{Brian: reflect on why people like it -- why all the features together make it work?}

% Mention the love-and-hate relationship about linearity

% Confirmation of the prevalent of out-of-order execution and single-cell risk management in prior work

\subsection{\new{Handling the }\old{Happy (and Sad) Side Effects of Clarity}}\label{subsec:discussion-clarity-side-effects}
Clarity is one quality attribute among many others that users value in their notebooks~\cite{huang2025howscientists}, notable others including correctness and reproducibility.
Although not our design intention, by promoting clarity in the program state (mainly out of concerns for user comprehension), \sys led to the happy side effects of supporting correctness and reproducibility---our sanity check on the 13 notebooks used in the study showed that they all reproduced the same results after the study (\autoref{subsec:system-usage}).

Maintaining clarity still comes with sad side effects.
First, although our main goal was to alleviate the well-known tension between clarity and explorability~\cite{Rule2018:Exploration, liuRefactoringComputationalNotebooks2023a, Raghunandan2023:Code, keryStoryNotebookExploratory2018, huang2025howscientists}, the tension remains in situations where clarity is not the main goal.
For example, clarity need not be maintained in using notebooks entirely for scratch work (P8, P9) or small tasks that might eventually be scripted (P9).
Second, there are also situations where other attributes like debuggability and efficiency take precedence over clarity.
In these scenarios, users might prefer nonlinear execution for quick debugging and avoiding re-processing large data (P2, P8, P9), where the enforced linearity can become restrictive.
% Perhaps the solution is to further embrace flexible support for clarity, as participants noted in suggestions for \sys (\autoref{subsec:results-usability}), via \emph{context-aware clarity adjustments}.

Perhaps the solution is to, \eg, offer a toggle back to ordinary Jupyter execution semantics, and to incorporate other \del{adjustments for}\new adjustable \old clarity support suggested by participants:
\new
\begin{enumerate}
\item Better visual layout of the \spad: When one has a lot of scratch work, having better auto-layout algorithms (P1, P7, P10) (\eg, like Enso~\cite{EnsoAnalytics}) could help better organize the \spad;
\item Placing pinned cells and scratch sections more freely depending on the notebook task: For more exploratory work, scratch sections need not be attached to any part of a notebook, so they could instead become resizable and floating windows that can be placed anywhere on the screen (P2, P5, P9, P10, P12) just like in StickyLand~\cite{Wang2022:StickyLand};
\item More integrated AI support that facilitates notebook clarity, particularly in documentation, such as documenting multiple cells (P7), deriving hypotheses from code (P2, P9), summarizing output (P2) and errors (P12), refactoring scratch work into a notebook cell (P3), and code generation (P6).
\end{enumerate}
\old

\new
Furthermore, future notebook systems could incorporate a ``slider'' that corresponds to a spectrum of support for clarity, with the most relaxing that optimizes for rapid exploration to the most restricted that prioritizes notebook interpretability, depending on user preferences and task requirements.
\old

% implications of promoting notebook clarity: correctness, reproducibility

% tensions between clarity and explorability and debuggability stil exist -- reuse the ``tensions'' paragraph

\subsection{\new{Broader Design Implications for Information Systems}\del{Towards Better Notebook Sharing}}\label{subsec:discussion-sharing}
\new{Our system and study results can shed light on future information systems that need to balance exploration and clarity, such as word processors and presentation programs.
Specifically, our three design goals directly inform the following general design goals for such information systems: \textbf{providing structural flexibility}, \textbf{enabling both ephemeral and evolving ideation}, and \textbf{aligning visual layout with internal logic}.}

Our study further revealed opportunities for \new{\textbf{supporting on-the-fly information sharing}.
Specifically for notebook sharing, this entails} \old teaching, collaborative editing, and output-driven sharing scenarios\new{ that require little overhead in cleaning}. \old 
For teaching, \sys already supports effort-free notebook cleaning, a big alleviation for instructors (P5, P11) who often have little time for cleaning up notebooks that become messy during lectures but then need to be shared with the class.
For collaborative editing, there could be better support for documentation, such as headings for scratch sections, similar to Janus~\cite{ruleAidingCollaborativeReuse2018} but automated like Themisto~\cite{wang2022documentation}, to facilitate understanding the \spad explorations from one other.
Finally, when clarity matters less for the code but more for the output (P6), we could incorporate functions such as output-only dashboards~\cite{Wang2022:StickyLand} or high-level notebook summaries.
%These scenarios are beyond \sys's existing support for the always-clear \emph{authoring} process.
\new{The above notebook sharing scenarios apply to general information systems and can benefit from similar support for always-clear authoring.}\old

\section{Conclusion}
\new
We proposed always-clear notebook authoring, a paradigm that
\old
that supports both clarity and exploration throughout the entire notebook lifecycle, instantiated in a prototype for Jupyter called \sys.
An exploratory study (N=13) showed that \sys features collectively improved the experience of maintaining clarity, 
\new
maintained support for realistic notebook tasks,
and
\old
enabled novel strategies for clarity.
Our study highlights the promise of \sys in supporting always-clear notebook authoring and 
reveals \new{key design opportunities for future notebook systems that provide configurable clarity support, and for general information systems that balance exploration and clarity.}\old

\begin{acks}
Our thanks to Savitha Ravi, Saketh Kasibatla, and the Foundation Interface Lab for helpful feedback on earlier prototypes.
This work was supported by U.S. National Science Foundation Grants No. 2107397 (\emph{Human-Centric Program Synthesis}) and No. 2432644 (\emph{Direct Manipulation for Everyday Programming}).
\end{acks}

%\end{sloppypar}

\bibliographystyle{ACM-Reference-Format}

%%%% bibliography starts here %%%%


%\bibliography{citations}%% Commented by merge tool

%%%% appendix.tex starts here %%%%

\appendix

\section{AI-Generated Cell Summary for Cell Folding in \sys}\label{app:folding}

To facilitate cell folding (\ie, showing a more meaningful information than ``$\cdots$'' as in JupyterLab), \sys generates a summary for every executed cell.
When a code cell is executed, \sys records the cell code, summarizes the code and variables newly defined with GPT-4o, and inserts the summary at its beginning.
\sys only re-summarizes if the cell code (with comments and whitespace removed) differs from prior record.
The prompt is a formatted string in TypeScript:

\begin{lstlisting}[
breakatwhitespace=true,
language=TypeScript,
postbreak=\mbox{\textcolor{gray}{$\hookrightarrow$}\space},
]
const priorCodePrompt = prevCode ? `The code prior to this cell that has been executed is:\n\n\`\`\`\n${prevCode}\`\`\`\n\n` : '';

const prompt =
`I am writing code in this notebook cell:

\`\`\`
${origCodeSansOutdatedComments}
\`\`\`

${priorCodePrompt}

Summarize the code in this cell in a comment with no more than ten words.

Summarize the variables newly defined in this cell in another comment with no more than ten words.

Return only the two comments starting with #, separated by a newline.`;
\end{lstlisting}

\noindent
\scode{origCodeSansOutdatedComments} is the code of the cell to be summarized with existing summaries removed, if any.
\scode{prevCode} is code from cells linearly above to the given cell that have been executed.
We only include \scode{prevCode} in the prompt (\ie, in \scode{priorCodePrompt}) if it is non-empty.

\end{document}